\documentclass[11pt]{article}
\usepackage{graphicx}
\usepackage{upgreek}
\usepackage{amsmath}
\usepackage{amssymb}
\usepackage{setspace}
\title{An extension to GUM methodology: degrees-of-freedom calculations for
correlated multidimensional estimates}
  \author{
    R. Willink\\\texttt{robin.willink@gmail.com}
    \and B. D. Hall\\
    Measurement Standards Laboratory of New Zealand,\\
    Callaghan Innovation,\\
    P.O. Box 31-310,
    Lower Hutt 5040,
    New Zealand\\
    \texttt{blair.hall@callaghaninnovation.govt.nz}
  }

\begin{document}
\maketitle

\begin{abstract}
The {\em Guide to the Expression of Uncertainty in Measurement} advocates the use of an `effective number of degrees of freedom' for the calculation of an interval of measurement uncertainty. However, it does not describe how this number is to be calculated when (i) the measurand is a vector quantity or (ii) when the errors in the estimates of the quantities defining the measurand (the `input quantities') are not incurred independently. An appropriate analysis for a vector-valued measurand has been described ({\em Metrologia} {\bf 39} (2002) 361--9), and a method for a one-dimensional measurand with dependent errors has also been given ({\em Metrologia} {\bf 44} (2007) 340--9). This paper builds on those analyses to present a method for the situation where the problem is multidimensional {\em and} involves correlated errors. The result is an explicit general procedure that reduces to simpler procedures where appropriate. The example studied is from the field of  radio-frequency metrology, where measured quantities are often complex-valued and can be regarded as vectors of two elements.
\end{abstract}

\section{Introduction}

The {\em Guide to the Expression of Uncertainty in Measurement} (the {\em Guide} or GUM) gives a practical procedure for the evaluation of measurement uncertainty. That procedure is based around the idea of adding individual estimates of variances to obtain an estimate of overall measurement variance and then combining the numbers of `degrees of freedom' for the individual estimates to obtain an effective number of degrees of freedom for this overall estimate. The {\em Guide}'s procedure is only applicable with one-dimensional quantities, which in this paper we will call `scalars', but there are a number of areas of metrology that must deal with multidimensional quantities also. Perhaps the most common involve complex-valued quantities, which arise in association with wave phenomena in fields such as acoustics, optics and electrical metrology. In particular, metrology at radio and microwave frequencies (RF) is often concerned with measuring reflection and transmission coefficients when electromagnetic waves propagate in a network of components. Standard RF instrumentation can be used to accumulate simultaneous measurements of several such coefficients to form a sample of data. When the measurand is related to a number of these coefficients, subsequent data processing will use the sample data to obtain the measurement result, and the associated sample variances and covariances should be used to evaluate the measurement uncertainty. This paper describes a method for calculating the number of effective degrees of freedom in such cases, so that a region of expanded uncertainty can be evaluated.

When the measurand is a scalar estimated by averaging $n$ replicated measurement results, the procedure of the {\em Guide} involves forming a confidence interval using the $t$-distribution with $n-1$ degrees of freedom, as in the classical statistical approach to estimating the mean of a normal distribution \cite[clause 4.2.3, note 1]{GUM}. The general problem addressed in the {\em Guide}, however, involves a measurand that is a function of many scalars, with this function being locally approximately linear. The uncertainty of measurement is then analogous to the uncertainty in the estimate of the weighted sum of many scalars. In that case there is no $t$-distribution from which we can construct an exact confidence interval, so the {\em Guide} advocates a method that uses a $t$-distribution but is approximate. This method uses the Welch-Satterthwaite (W-S) formula to calculate the best number of degrees of freedom \cite[appendix G.4]{GUM}. The W-S formula is derived and justified under the classical, i.e.\ frequentist, approach to statistical analysis.  It is applicable when the quantity of interest is the sum of scalars each independently estimated by averaging measurement results \cite{Welch37,Welch47,Satterthwaite}. Therefore, the use of the W-S formula in metrology is most appropriate when the classical approach is adopted and when the dominant components of uncertainty are evaluated by Type A methods. That is the primary context for extension of the {\em Guide} methodology by generalization of the W-S formula. Our analysis is firmly grounded within the classical statistical framework,
in which the W-S formula applies. Working in a non-classical context would involve a different formula as the starting point \cite[appendix]{doe}.

\subsection{Notation and terminology}

As in the {\em Guide}, the upper-case letters $X$ and $Y$ denote unique unknown values of fixed scalar quantities and the lower-case letters $x$ and $y$ indicate known estimates of these quantities. Similarly, upper-case bold letters like ${\pmb X}$ and ${\pmb Y}$ and lower-case bold letters like ${\pmb x}$ and ${\pmb y}$ denote fixed vector quantities and their estimates. Upright bold symbols such as ${\bf u},{\boldsymbol \Sigma}$ and ${\bf s}$ are used for matrices. In classical statistics an observation, say $x$, is generally seen as the outcome of a random variable, and this random variable will often be denoted by the corresponding capital letter. However, because the symbol $X$ is being used for a different purpose in this paper, the random variable of which $x$ is the outcome will be denoted by $\check{x}$ \cite{generalize}. Physicists tend to think in terms of the data,
not the random variables that gave rise to the data, so there is merit in minimizing explicit reference to random variables. Thus instead of writing
$\check{x}\sim {\rm N}(0,4)$, for example, to mean that the random variable $\check{x}$ {\em possesses} the normal distribution with mean 0 and variance 4, we will write
\[
x\leftarrow {\rm N}(0,4)
\]
to mean that to the number $x$ {\em was drawn from} the normal distribution with mean 0 and variance 4 \cite{MUAP}.\footnote{This usage of the symbol $\leftarrow$ should not be confused with the usage of Wang, Hannig and Iyer \cite{pivot}.} Similarly, instead of writing $\sigma^2(\check{x})$ or ${\rm var}\,\check{x}$ to mean the variance of the random variable $\check{x}$ we will write $\sigma^2_{\rm p}(x)$ or ${\rm var}_{\rm p} x$ to describe the same quantity as the {\em parent} variance of $x$.

The term `measurement result' is often used in metrology to represent the outcome of the measurement process. The basic context of the {\em Guide}, and the context of this paper, is where the measurand has an essentially unique unknown value that is estimated or approximated in the measurement. So, in keeping with common statistical terminology, we will often describe a measurement result as an `estimate'.

\subsection{A basic result: estimating a multidimensional quantity from repeated measurement}
\label{basicsec}

The foundation of the methodology to be discussed in this paper is the following result in multivariate statistical analysis. Suppose that $n$ measurements are made of an unknown ${q}$-dimensional quantity ${\pmb X}$ and that the results are ${\pmb x}_{[1]},\ldots,{\pmb x}_{[n]}$. The sample mean vector of these results is defined to be
\begin{eqnarray*}
{\bar {\pmb{x}}}\;&\equiv& {\frac{1}{n}\sum_{k=1}^n
{\pmb x}_{[k]}}
\end{eqnarray*}
and the sample covariance matrix of these results is defined to be the ${q}\times{q}$ matrix
\begin{eqnarray*}
{\bf s}&\equiv&
{\frac{1}{n-1}\sum_{k=1}^n \left({\pmb
x}_{[k]}-{\bar {\pmb{x}}}\right)\left({\pmb
x}_{[k]}-{\bar {\pmb{x}}}\right)^\prime}.
\end{eqnarray*}
(More strictly, this could be called the sample estimate of the covariance matrix.) The best point estimate of ${\pmb X}$ available from this data is ${\bar {\pmb{x}}}$, so we make the definition ${\pmb{x}}\equiv \bar{\pmb{x}}$ in order to let ${\pmb{x}}$ symbolize this estimate. Suppose that each ${\pmb x}_{[j]}$ is independently drawn from the ${q}$-dimensional multivariate normal distribution with mean vector ${\pmb X}$ and some covariance matrix ${\boldsymbol\Sigma}$, which is a distribution denoted ${\rm N}_{q}({\pmb X},{\boldsymbol\Sigma})$. That is, suppose
${\pmb x}_{[j]} \leftarrow {\rm N}_{q}({\pmb X},{\boldsymbol\Sigma})$. Then ${\pmb x}$ is drawn from the multivariate normal distribution with mean vector ${\pmb X}$ and covariance matrix ${\boldsymbol\Sigma}/n$, while $(n-1){\bf s}$ is independently drawn from the Wishart distribution with scale matrix ${\boldsymbol\Sigma}$ and $n-1$ degrees of freedom, which we denote by ${\rm W}_{q}(n-1,{\boldsymbol\Sigma})$ \cite{ChatfieldCollins}. That is,
\begin{eqnarray}
{{\pmb{x}}}&\leftarrow &{\rm N}_{q}({\pmb X},{\boldsymbol\Sigma}/n)\nonumber\\
(n-1){\bf s}&\leftarrow& {\rm W}_{q}(n-1,{\boldsymbol\Sigma})\nonumber,
\end{eqnarray}
with ${{\pmb{x}}}$ and ${{\bf{s}}}$ being drawn independently.
It follows that
\begin{equation}
({\pmb{x}}-{\pmb X})^\prime ({\bf s}/n)^{-1} ({\pmb{x}}-{\pmb X}) \leftarrow {\rm T}^2_{n-1,{q}}
\label{Hotelling}
\end{equation}
where ${\rm T}^2_{n-1,{q}}$ signifies the distribution of Hotelling's $T^2$ with $n-1$ degrees of freedom and second parameter ${q}$ \cite{ChatfieldCollins}.

The distribution ${\rm T}^2_{\nu,{q}}$ is the distribution of $\nu{q}(\nu+1-{q})^{-1}F_{{q},\nu+1-{q}}$ where $F_{{q},\nu+1-{q}}$ is a random variable with the $F$-distribution with ${q}$ degrees of freedom in the numerator and $\nu+1-{q}\,$ degrees of freedom in the denominator. Let $F_{\nu_1,\nu_2,0.95}$ denote the $0.95$-quantile of the distribution of $F_{\nu_1,\nu_2}$. Then, from (\ref{Hotelling}), the ellipsoidal region of vectors ${\pmb X}^\ast$ described by the equation
\[
({\pmb{x}}-{\pmb X}^\ast)^\prime ({\bf s}/n)^{-1} ({\pmb{x}}-{\pmb X}^\ast)< \frac{(n-1)r}{n-{q}}F_{{q},n-{q},0.95},
\]
is the realization (or outcome) of a 95\% confidence region for ${\pmb X}$. In other words, whatever the values of the unknown parameters ${\pmb X}$ and ${\boldsymbol\Sigma}$, the overall procedure had probability 0.95 of generating a region containing ${\pmb X}$. This procedure includes the steps of choosing $n$ without reference to the data, making the measurements and calculating the ellipsoid.

The result just described is a standard result in multivariate statistics. The subject of this paper is the generalization of this result for situations of metrological interest. This process begins with a statement of the basic mathematical structure of a multidimensional measurement problem. This statement is analogous to the description of a one-dimensional problem found in the {\em Guide}.

\section{The {\em Guide} formulation applied to a multidimensional problem}

The {\em Guide} addresses the problem where the unique unknown value $Y$ of a scalar measurand is related by a known function to the unique unknown values $X_1,\ldots,X_m$ of $m$ scalar input quantities. In generalizing this formulation to the multidimensional situation, it seems sensible to adopt analogous notation. So we consider the measurement of a quantity ${\pmb Y}$ of dimension $p$ that is related by a known function $f$ to the unique unknown values ${\pmb X}_1,\ldots,{\pmb X}_m$ of $m$ input quantities of dimension\footnote{Situations where $q>p$ are not uncommon in RF metrology. For example, when the measurand is an attenuation or power, which are real-valued, the measurement equation will include complex influence quantities related to the scattering of electromagnetic waves by the measurement network.}  ${q}\ge p$. (If necessary, input quantities can be padded with dummy components to bring them all to the same dimension, ${q}$.) The vector value of the measurand is ${\pmb Y}=f({\pmb X}_1,\ldots,{\pmb X}_m)$ and the result obtained is ${\pmb y}=f({\pmb x}_1,\ldots,{\pmb x}_m)$. The familiar `linear approximation' to the measurement error written for the multidimensional case is
\begin{equation}
{\pmb y}- {\pmb Y}\approx \sum_{i=1}^m {\bf c}_i ({\pmb x}_i-{\pmb X}_i)
\label{linapprox}
\end{equation}
where
\begin{equation}
{\bf c}_i \equiv \left. \frac{\partial f({\pmb x}_1,\ldots,{\pmb x}_{i-1},{\pmb z},{\pmb x}_{i+1},\ldots,
{\pmb x}_m)}{\partial {\pmb z}}\right|_{{\pmb z}={\pmb x}_i}.
\label{cimatrix1}
\end{equation}
Here ${\pmb y}$ and ${\pmb Y}$ are column vectors of length $p$, ${\pmb x}_i$ and ${\pmb X}_i$ are column vectors of length ${q}$, and ${\bf c}_i$ is a $p\times {q}$ matrix of sensitivity coefficients. The quantity $\partial {\pmb {f}}/\partial {\pmb z}$ for a $p$-dimensional vector ${\pmb {f}}\equiv ({f}_{1},\ldots,{f}_{p})^\prime$ and a ${q}$-dimensional vector ${\pmb z}\equiv (z_{1},\ldots,z_{q})^\prime$ is defined to be the $p\times {q}$ matrix
\begin{equation}
\frac{\partial {\pmb {f}}}{\partial {\pmb z}} \equiv
\left[
\begin{array}{ccc}
\frac{\partial {f}_{1}}{\partial z_1}&\cdots&\frac{\partial {f}_{1}}{\partial z_{q}}\\
\vdots&\ddots&\vdots\\
\frac{\partial {f}_{p}}{\partial z_1}&\cdots&\frac{\partial {f}_{p}}{\partial z_{q}}
\end{array}
\right].
\label{cimatrix2}
\end{equation}

The ${\pmb x}_i$ quantities might be two-dimensional vectors representing complex numbers, with the real component being the first element. If so and if the function $f$ is complex and differentiable, the $2 \times 2$ matrices ${\bf c}_i$ may be obtained with less effort using the Cauchy-Riemann equations \cite{hall04}. In that case, we evaluate the complex partial derivative of $f({\pmb x}_1,\ldots,{\pmb x}_m)$ with respect to ${\pmb x}_i$ and obtain the number $a + b\,\mathrm{j}$, say, and then we can write
\begin{equation}
{\bf c}_i = \left[
\begin{array}{rr}
a & -b \\
b & a
\end{array}
\right].
\label{cimatrixcomplex}
\end{equation}

The law of propagation of uncertainty for scalar quantities given in the {\em Guide} \cite[clause 5.1.2]{GUM} can be extended to the vector case.
Let ${\bf v}_{\pmb y}$ be the $p\times p$ covariance matrix to be associated with ${\pmb y}$. When each ${\pmb X}_i$ is measured independently, the generalization of the law of propagation of uncertainty is ${\bf v}_{\pmb y} = \sum_{i=1}^m {\bf c}_i {\bf v}_i {\bf c}_i^\prime$ where ${\bf v}_i$ is the ${q}\times {q}$ covariance matrix associated with ${\pmb x}_i$. For the kind of uncertainty evaluation described in this paper,
\[
{\bf v}_i = \frac{{\bf s}_i}{n_i}.
\]
For the situation where the ${\pmb X}_i$ quantities are not measured independently, we can write (\ref{linapprox}) in the form
\begin{equation}
{\pmb y}-{\pmb Y}\approx {\bf c}({\pmb x}-{\pmb X})
\label{linapprox2}
\end{equation}
where ${\pmb x}=({\pmb x}_1^\prime,\ldots,{\pmb x}_m^\prime)^\prime$, $\;{\pmb X}=({\pmb X}_1^\prime,\ldots,{\pmb X}_m^\prime)^\prime$ and ${\bf c}$ is the appropriate $p\times m{q}$ matrix. Then the law becomes ${\bf v}_{\pmb y} = {\bf c} {\bf v}_{\pmb x} {\bf c}^\prime$ for the appropriate matrix ${\bf v}_{\pmb x}$.

\subsection{Calculation of a confidence region}

The law of propagation of uncertainty can be used to calculate a $p$-dimensional region of measurement uncertainty when the measurement error is well modelled as having arisen from the normal distribution with mean zero and known $p\times p$ covariance matrix. For example, the calculation of a confidence region for ${\pmb Y}$ in (\ref{linapprox}) is straightforward when each ${\pmb X}_i$ is independently estimated from a sample of size $n_i$ as in Section~\ref{basicsec} and when the corresponding covariance matrix ${\bf \Sigma}_i$ is known. Then the realization of a 95\% confidence region for ${\pmb Y}$ is the region of vectors ${\pmb Y}^\ast$ described by
\begin{equation}
({\pmb{y}}-{\pmb Y}^\ast)^\prime {\bf v}_{\pmb y}^{-1} ({\pmb{y}}-{\pmb Y}^\ast)\;<\; \chi^2_{p,0.95}
\label{chisqregion}
\end{equation}
where ${\bf v}_{\pmb y} = \sum_{i=1}^m {\bf c}_i {\bf \Sigma}_i{\bf c}_i^\prime/n_i$ and $\chi^2_{p,0.95}$ is the 95th percentile of the chi-square distribution with $p$ degrees of freedom. Similarly, when (\ref{linapprox2}) is a useful representation this result holds with ${\bf v}_{\pmb y} = {\bf c} {\bf \Sigma_{\pmb x}} {\bf c}^\prime$ for the appropriate covariance matrix ${\bf \Sigma_{\pmb x}}$.

However, ${\bf \Sigma}_{i}$ will usually be unknown and so any matrix ${\bf v}_{\pmb y}$ calculated for use in (\ref{chisqregion}) must be seen as an inexact estimate of the appropriate parent covariance matrix. Consequently,  (\ref{chisqregion}) will not necessarily give uncertainty regions that enclose the measurand with adequate frequency. In this paper we address this problem by giving a degrees-of-freedom formula for use with Hotelling's distribution analogous to the W-S formula used with a $t$-distribution in the univariate case. With $\nu_{\rm eff}$ being the effective number of degrees of freedom obtained using this formula, the region of expanded uncertainty is described by the inequality
\begin{equation}
({\pmb{y}}-{\pmb Y}^\ast)^\prime {\bf v}_{\pmb y}^{-1} ({\pmb{y}}-{\pmb Y}^\ast)< \frac{\nu_{\rm eff}{q}}{\nu_{\rm eff}+1-{q}}F_{{q},\nu_{\rm eff}+1-{q},0.95}
\label{basicregion}
\end{equation}
for some appropriate matrix ${\bf v}_{\pmb y}$. (The region given by (\ref{chisqregion}) is obtained as $\nu_{\rm eff}\rightarrow\infty$.) Section~\ref{newsec1} reviews and extends a formula applicable with independent components of error. Section~\ref{newsec2} gives the main advancement of this paper, which is a more general formula that allows a common form of dependency between components.

\section{Finite degrees of freedom: independent estimates}
\label{newsec1}

Suppose that a $p$-dimensional measurement result ${\pmb y}$ has been formed by applying the function $f$ to independent measurements of
the $q$-dimensional input quantities, and that the $m$ parent covariance matrices of the errors in these measurements are unknown. We look for a $p$-dimensional uncertainty region described by (\ref{basicregion}) for a dummy vector ${\pmb Y}^\ast$, for some appropriate matrix ${\bf v}_{\pmb y}$ and for some effective number of degrees of freedom $\nu_{\rm eff}$. Willink and Hall \cite{classical} have discussed this problem in the context of estimating the sum of $m$ multivariate means $\boldsymbol{\mu} \equiv \sum \boldsymbol{\mu}_i$. If we identify the vector $\boldsymbol{\mu}_i$ with ${\bf c}_i{\pmb X}_i$ for some constant matrix $c_i$ then their results can be applied in the situation described by the equation ${\pmb y}-{\pmb Y}\approx \sum_{i=1}^m {\bf c}_i({\pmb x}_i-{\pmb X}_i)$ where $p={q}$. In that case, ${\bf v}_{\pmb y}=\sum_{i=1}^m {\bf w}_i$ with the matrix ${\bf w}_i\equiv (w_{i\cdot jk})$ being equal to ${\bf c}_i  {\bf v}_i {\bf c}_i^\prime$. Willink and Hall examined a number of possibilities for the corresponding effective number of degrees of freedom. The choice now strongly favoured,\footnote{The other independent option studied was denoted $\nu_{\rm gv}$ \cite[eq. 13]{classical}. There are several reasons for favouring the use of $\nu_{\rm tv}$. These are: (i) $\nu_{\rm tv}$ satisfies the requirement of `transferability' in the overall evaluation of measurement uncertainty \cite[clause 0.4]{GUM} \cite[sec.~5]{classical}, (ii) $\nu_{\rm tv}$ is more conservative \cite[sec. 4]{classical} and (iii) $\nu_{\rm gv}$ is undefined when a $p$-dimensional problem is treated as if it were a problem in a greater number of dimensions, which should theoretically be possible. Each $p$-dimensional quantity can be represented as a quantity in $p+k$ dimensions with the last $k$ components being zero. The determinants that define $\nu_{\rm gv}$ both become zero, and the result would be $0/0$.} which was the option denoted $\nu_{\rm tv}$ \cite[eqs. 12,14]{classical}, is
\begin{equation}
\nu_{\rm eff}=
\frac{\displaystyle \sum_{j=1}^p\sum_{k=j}^p \left\{\sum_{i} w_{i\cdot jj}\sum_{i} w_{i\cdot kk}+\left(\sum_{i} w_{i\cdot
jk}\right)^2\right\}}{\displaystyle \sum_{j=1}^p\sum_{k=j}^p\left\{\sum_{i}\left(w_{i\cdot jj}w_{i\cdot kk}+w_{i\cdot jk}^2\right)/(n_i-1)\right\}}.
\label{nueff14}
\end{equation}
For this application, the summation over $i$ is from $i=1$ to $i=m$. When $m=1$ we obtain $\nu_{\rm eff}=n_1-1$.

The method of the earlier paper \cite{classical} is now presented in a way that explicitly shows its applicability with (\ref{linapprox}) and which provides a basis for the analysis of Section~\ref{newsec2}. We employ a result from the theory of multivariate analysis \cite[Theorem 3.3.11]{GuptaNagar}.
\begin{quote}
If ${\bf {v}}\leftarrow {\rm W}_{q}(\nu,{\boldsymbol\Omega})$ and
    ${\bf c}$ is a constant matrix of dimension $p\times {q}$ and rank $p\le {q}$ then ${\bf c}\,{\bf {v}}\, {\bf c}^\prime
\leftarrow {\rm W}_p(\nu,{\bf c}\,{\boldsymbol\Omega}\, {\bf c}^\prime)$.\\[-9mm]
\begin{equation}\label{100}\end{equation}
\end{quote}
Thus, from the relationship $(n_i-1){\bf s}_i \leftarrow {\rm W}_{q}(n_i-1,{\boldsymbol \Sigma}_i)$, (which is seen above and is also seen in expression 2 of Willink and Hall \cite{classical}), we obtain the result $(n_i-1){\bf c}_i{\bf s}_i{\bf c}_i^\prime \leftarrow {\rm W}_p(n_i-1,{\bf c}_i{\boldsymbol \Sigma}_i{\bf c}_i^\prime)$. Subsequently, expressions 6 and 7 of Willink and Hall \cite{classical} become
\[
{\pmb y}-{\pmb Y} \stackrel{\rm app}{\leftarrow} {\rm N}_p\left({\bf 0},  \sum_{i=1}^m {\bf c}_i{\bf \Sigma}_i{\bf c}_i^\prime/n_i\right)
\]
and
\[
\sum_{i=1}^m {\bf c}_i{\bf v}_i{\bf c}_i^\prime  \stackrel{\rm app}{\leftarrow}
\frac{b}{\nu} {\rm W}_p\left(\nu,  \sum_{i=1}^m {\bf c}_i{\bf \Sigma}_i{\bf c}_i^\prime/n_i \right)
\]
where the scalars $b$ and $\nu$ are to be determined. (Here $\stackrel{\rm app}{\leftarrow}$ means `is drawn from a distribution approximated by' and
${\bf 0}$ denotes a vector of zeroes.) The analysis then proceeds as in the earlier paper \cite{classical} but with the elements of ${\bf c}_i{\bf v}_i{\bf c}_i^\prime$ replacing the corresponding elements of ${\bf v}_i$, i.e. ${\bf s}_i/n$. This implies that an appropriate figure for the effective number of degrees of freedom when the measurement is described by (\ref{linapprox}) is given by (\ref{nueff14}) with
\begin{equation}
{\bf w}_i = {\bf c}_i  {\bf v}_i {\bf c}_i^\prime.
\label{wcuc}
\end{equation}
Subsequently, the realised confidence region contains the vectors ${\pmb Y}^\ast$ that satisfy (\ref{basicregion}) with
\begin{equation}
{\bf v}_{\pmb y} = \sum_{i=1}^m  {\bf w}_i.
\label{vy12}
\end{equation}

The bivariate case of $p=2$ is of particular interest because of the importance of complex quantities. In that case, (\ref{nueff14}) simplifies to
\begin{equation}
\nu_{\rm eff}=\frac{
2\left(\sum_{i} w_{i\cdot 1{}1}\right)^2+
\sum_{i} w_{i\cdot 1{}1}\sum_{i} w_{i\cdot 2{}2}+\left(\sum_{i} w_{i\cdot 1{}2}\right)^2+
2\left(\sum_{i} w_{i\cdot 2{}2}\right)^2}{
\sum_{i} \left(2 w_{i\cdot 1{}1}^2 + w_{i\cdot 1{}1}w_{i\cdot 2{}2}+w_{i\cdot 1{}2}^2+2 w_{i\cdot 2{}2}^2\right)/(n_i-1)
}
\label{ADFadf}
\end{equation}
with summation being from $i=1$ to $i=m$ \cite[Appendix 1]{classical}.

\subsection{Example 1}
\label{example1sec}

In this section we present the first of two examples contained in this paper. Code for their analysis is provided in the appendix. In these examples, we do not require that bold notation always be used for multidimensional quantities.

\begin{figure}[ht]
\begin{center}
\vskip 1cm
\includegraphics[width=11cm]{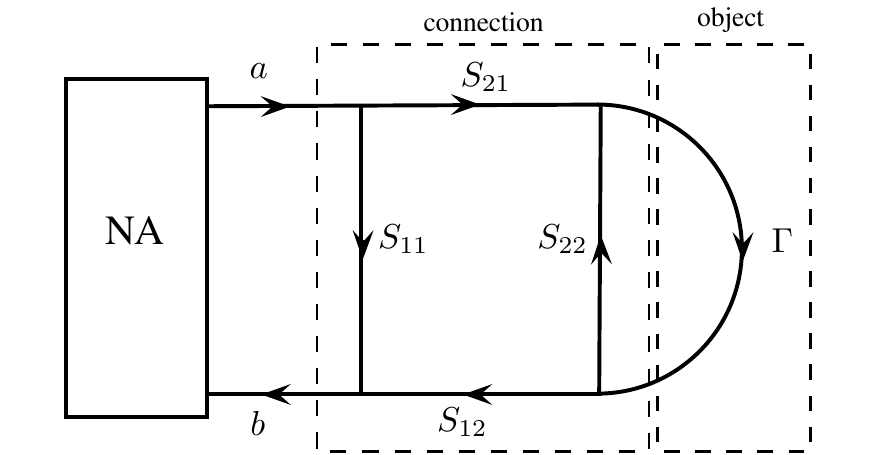}
\vskip 1cm
\caption{A signal flow diagram showing the measurement of the reflection coefficient $\Gamma$ of an object. A network analyser (NA) is connected to the component of interest via a network that is characterised by four complex $S$-parameters. The complex ratio $\Gamma^\prime = b/a$, of reflected to incident amplitudes, is recorded by the analyser so $\Gamma$ can be estimated when there is additional information available about the $S$-parameters.}
\label{fig_two_port_graph}
\end{center}
\end{figure}

The measurement of the reflection coefficient $\Gamma$ of a physical object is common in RF metrology. Often $\Gamma$ cannot be directly observed, and instead the quantity measured in the experiment is
\begin{equation}
\Gamma^\prime = S_{11} + \frac{S_{12}S_{21}\Gamma}{1 - S_{22}\Gamma}
\label{gammaprime}
\end{equation}
where $S_{11}, S_{12}, S_{21}$ and $S_{22}$ are the $S$-parameters of a network used to connect the object of interest to the measuring instrument \cite{pozar}, as shown schematically in Fig~\ref{fig_two_port_graph}. The complex ratio of reflected and incident wave amplitudes seen at the instrument is recorded. In general, the interconnecting network has non-ideal transmission characteristics, which are represented by $S_{21}$ in the forward direction and $S_{12}$ in the reverse direction, and also it may involve reflections at either end, as represented  by $S_{11}$ and $S_{22}$. Each $S$-parameter is complex, so all six quantities featuring in (\ref{gammaprime}) are complex.

The measurand $\Gamma$ is expressible as
\begin{equation}
\Gamma = \frac{{\Gamma}^\prime - {S}_{11}}{{S}_{12}{S}_{21} + {S}_{22}({\Gamma}^\prime - {S}_{11})}.
\label{Gamma}
\end{equation}
A common assumption is that the denominator in (\ref{Gamma}) is approximately $1 + 0\mathrm{i}$, in which case the measurement result becomes
\begin{equation}
\hat{\Gamma} = \hat{\Gamma}^\prime - \hat{S}_{11}.
\label{hatGamma1}
\end{equation}
The figures $\hat{\Gamma}^\prime$ and $\hat{S}_{11}$, which are the estimates of ${\Gamma}^\prime$ and ${S}_{11}$, are obtained independently. In this example, we consider the measurement of $\Gamma$ according to (\ref{hatGamma1}). Here $m=2$ and $p=q=2$. We identify the complex number ${S}_{11}$ with the two-dimensional vector ${\pmb X}_1$ and ${\Gamma}^\prime$ with the two-dimensional vector ${\pmb X}_2$ and, with a slight abuse of mathematical language, write ${\pmb X}_1={S}_{11}$ and ${\pmb X}_2={\Gamma}^\prime$.

Suppose that ${S}_{11}$ is estimated by averaging $n_1=5$ measurements $0.0242-0.0101\,\mathrm{j}$, $-0.0023+0.2229\,\mathrm{j}$, $0.0599+0.0601\,\mathrm{j}$, $0.0433+0.2100\,\mathrm{j}$ and $-0.0026+0.0627\,\mathrm{j}$. Suppose also that ${\Gamma}^\prime$ is estimated by averaging $n_2=5$ measurements $0.1648-0.0250\,\mathrm{j}$, $0.1568-0.0179\,\mathrm{j}$, $0.1598-0.1367\,\mathrm{j}$, $0.1198-0.0045\,\mathrm{j}$ and $0.3162-0.1310\,\mathrm{j}$. The estimate of ${S}_{11}$ is the corresponding sample mean ${\pmb x}_1 = \hat{S}_{11}= 0.02450+0.10912\,\mathrm{j}$. Likewise, the estimate of $\Gamma^\prime$ is the sample mean ${\pmb x}_2 =\hat{\Gamma}^\prime= 0.18348-0.06302\,\mathrm{j}$. From (\ref{hatGamma1}) we obtain ${\pmb y}=\hat{\Gamma} =  0.15898 - 0.17214\,\mathrm{j}$.

\begin{figure}[ht]
\vskip -5 cm
\hspace{-1cm}\includegraphics[width=16cm]{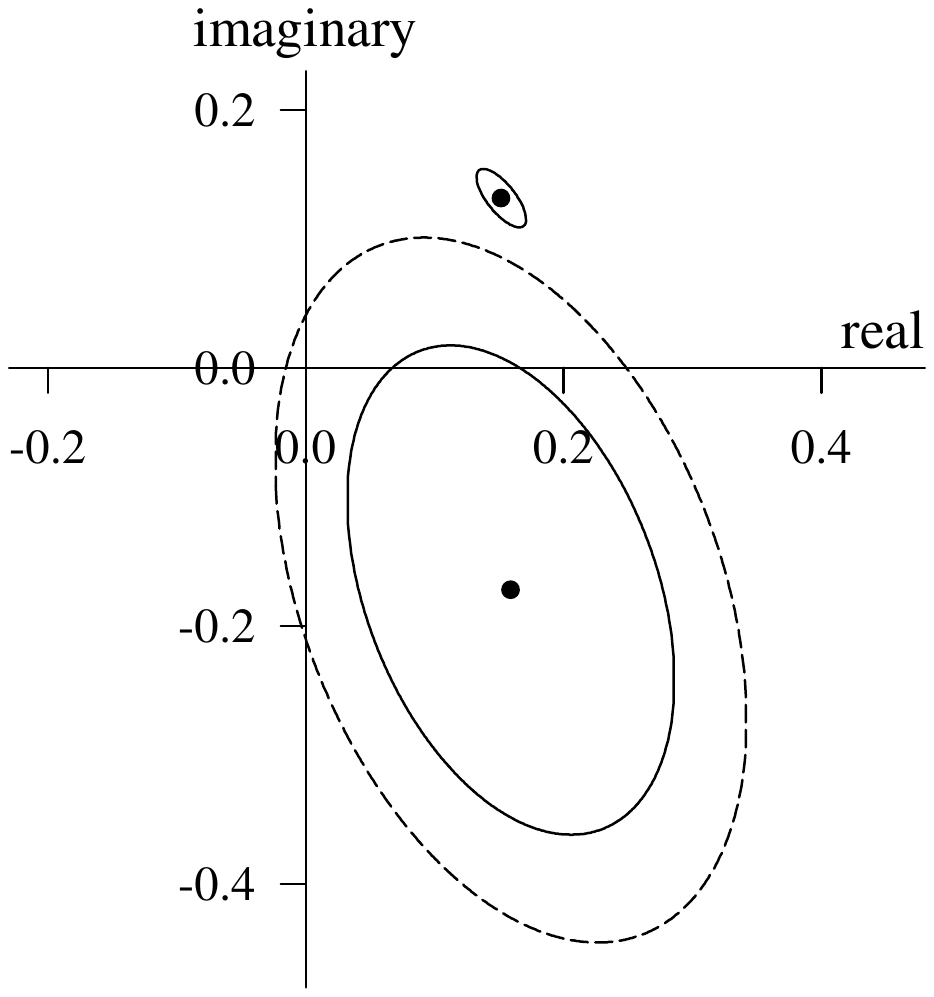}
\vskip -6.5cm
\caption{Measurement results (marked) and elliptical 95\% uncertainty regions for complex quantities. Large solid curve: the region formed from (\ref{basicregion}) with (\ref{ADFadf}) for Example 1. Dashed curve: the region obtained using the principles of {\em Supplement 2} \cite{supp2} in Example~1. Small solid curve: the region formed from (\ref{basicregion}) with (\ref{ADFadf}) for Example 2.}
\label{ellipses}
\end{figure}

The sample covariance matrices for the mean vectors ${\pmb x}_1$ and  ${\pmb x}_2$ are
\[
{\bf v}_1=
\left[
\begin{array}{rr}
0.1530  &-0.0797\\
-0.0797 &2.0947\\
\end{array}
\right] \times 10^{-3}
\hspace{1cm}{\rm and}\hspace{1cm}
{\bf v}_2=
\left[
\begin{array}{rr}
 1.1646 & -0.6459\\
-0.6459 & 0.8478\\
\end{array}
\right] \times 10^{-3}.
\]
From (\ref{cimatrixcomplex}) we find that
\[
{\bf c}_1=
\left[
\begin{array}{rr}
1 & 0 \\
0 & 1
\end{array}
\right]
\hspace{1cm}{\rm and}\hspace{1cm}
{\bf c}_2=
\left[
\begin{array}{rr}
-1 & 0 \\
0 & -1
\end{array}
\right]
\]
and then from (\ref{wcuc}) we find that ${\bf w}_1={\bf v}_1$ and ${\bf w}_2={\bf v}_2$. So, from (\ref{vy12}),
\[
{\bf v}_{\pmb y}=
\left[
\begin{array}{rr}
1.3175& -0.7256\\
-0.7256&2.9425
\end{array}
\right]\times 10^{-3}.
\]
Equation (\ref{ADFadf}) gives $\nu_{\rm eff}=6.85$ so the critical value on the right-hand side of (\ref{basicregion}) is 12.22. The 95\% region of uncertainty for $\Gamma$ is then the interior of the solid ellipse given in Figure~\ref{ellipses}.

\vskip 3mm
\noindent
{\em Comparison with the solution given in Supplement 2 to the Guide}\\[2mm]
\noindent
Our analysis may be contrasted with an analysis carried out in accordance with the second supplement to the {\em Guide} \cite{supp2} ({\em Supplement~2}).\footnote{The first two supplements to the {\em Guide} \cite{supp1,supp2} adopt a Bayesian-like view of analysis in assigning probability distributions to fixed unknown quantities \cite{MSTpaper}. If a Bayesian approach to uncertainty analysis had been intended at the time the {\em Guide} was written then the W-S formula would not have been theoretically appropriate.  So to adopt the approach taken in those supplements is to remove the theoretical basis for the use of the W-S formula. It is also to redefine the goal of uncertainty analysis from the concept of `success rate' in measurement to a concept of achievement that remains unspecified in the Bayesian metrology literature.} {\em Supplement~2} recommends that, after the observations have been made, the ${q}$-dimensional quantity ${\pmb X}_i$ is to be treated in the uncertainty evaluation as if it were a random vector with probability density function \cite[clause 5.3.2.2]{supp2}
\begin{equation}
g(\pmb{\upxi}) \propto \left\{1+\frac{1}{n_i-{q}}(\pmb{\upxi}-{\pmb{x}}_i)^\prime \left(\frac{{\bf \tilde{s}}_i}{n_i}\right)^{-1}(\pmb{\upxi}-{\pmb{x}_i})\right\}^{-n_i/2}
\label{posterior}
\end{equation}
where ${\bf \tilde{s}}_i=(n_i-1){\bf s}_i/(n_i-r)$. Here $\pmb{\upxi}$ is a dummy vector of length ${q}$.
This means that ${\pmb X}_i$ is attributed mean vector ${\pmb {x}}_i$ and covariance matrix \cite[clause 5.3.2.3]{supp2}
\begin{equation}
\frac{n_i-{q}}{n_i-{q}-2} \cdot\frac{{\bf \tilde{s}}_i}{n_i} \;\;=\;\;\frac{n_i-1}{n_i-{q}-2} \;{\bf {v}}_i.
\label{bigcov}
\end{equation}

{\em Supplement~2} describes how samples can be obtained from this distribution and, by implication, obtained from the distribution attributed to the measurand ${\pmb Y}$, which is $\Gamma$ \cite[clause 5.3.2.4]{supp2}. It also describes three ways of using the results to construct a `coverage region' said to contain ${\pmb Y}$ with a specified probability $P$ \cite[clause 7.7]{supp2}. We take $P=0.95$ and use a method equivalent to that described in clause 7.7.2 of {\em Supplement~2} to construct an elliptical coverage region for comparison with our result above.

In this example, ${q}=p=2$. The mean vector and covariance matrix of the distribution attributed to ${\pmb Y}$ are ${\pmb y}$ and
\begin{eqnarray*}
{\bf U}_{\pmb y} &=& \frac{n_1-1}{n_1-4}\;{\bf {v}}_1 + \frac{n_2-1}{n_2-4}\;{\bf {v}}_2\\[2mm]
&=&4{\bf v}_{\pmb y}.
\end{eqnarray*}
It was then found that 95\% (and no more) of the general values ${\pmb y}_{\rm MC}$ generated from the distribution attributed to ${\pmb Y}$ satisfied
\[
({\pmb{y}_{\rm MC}}-{\pmb y})^\prime {\bf U}_{\pmb y}^{-1} ({\pmb{y}}_{\rm MC}-{\pmb y})< 6.34.
\]
So the elliptical 95\% region for the measurand ${\pmb Y}=\Gamma$ is the region of points that satisfy this inequality. This region is indicated by the dashed curve on Figure~\ref{ellipses}. In terms of area, it is just over twice as large as our calculated confidence region.\footnote{Our context of multidimensional measurement reveals another difficulty with the recommendations of the supplements. {\em Supplement~2} advocates that a vector quantity estimated from $n$ measurements be assigned the probability density function (\ref{posterior}). The marginal distribution of any one element of that vector is found to be a shifted and scaled $t$-distribution with $n-2$ degrees of freedom \cite[section 3 and equation 3.7]{Geisser}.
However, {\em Supplement~1} advocates that a scalar quantity estimated from $n$ measurements be assigned a shifted and scaled $t$-distribution
with $n-1$ degrees of freedom \cite[clause 6.4.9.2 ]{supp1}. This inconsistency needs acknowledgement and explanation.}

In effect, this example involves the addition of two complex quantities. Hall \cite{hall06} has studied the success rates of different methods of evaluation in such a problem. In agreement with the result just given, he reported that the {\em Supplement 2} method (referred to there as the `H distribution') was conservative compared to the method of Willink and Hall \cite{classical}.

\section{Finite degrees of freedom: dependent estimates}
\label{newsec2}

The theory given in Section~\ref{newsec1} can be extended to permit dependence between the error processes that give rise to the estimates $\{{\pmb x}_i\}$. This is the main advancement made in this paper.

\vskip 3mm
\noindent
{\em Recalling the univariate case}\\[2mm]
\noindent
The problem of allowing dependence between components with finite degrees of freedom in the one-dimensional situation was considered by Willink~\cite{generalize}, who presented a generalization of the W-S formula for use with dependent estimates. A key relationship used in that analysis was as follows \cite[corollary 3.3.11.1]{GuptaNagar}:
\begin{quote}
If ${\bf {v}}\leftarrow {\rm W}_p(\nu,{\boldsymbol\Omega})$ and ${\pmb b}$ is a non-null column vector of length $p$ then $({\pmb b}^\prime\,{\bf {v}}\, {\pmb b})/({\pmb b}^\prime\,{\boldsymbol\Omega}\, {\pmb b})\leftarrow \chi^2_\nu$,
\end{quote}
where $\chi^2_\nu$ denotes the chi-square distribution with $\nu$ degrees of freedom. This result can be seen to be a special case of (\ref{100}). It enables us to make the following statement for scalar problems:
\begin{quote}
If two or more input quantities are estimated from a set of $n$ independent repeated simultaneous observations then the corresponding components of uncertainty should be combined into a single component attributed $n-1$ degrees of freedom.\\[-9mm]
\begin{equation}\label{300}\end{equation}
\end{quote}
This conclusion was described in Section~4.1 of that paper \cite{generalize} and was presented there as Result~1. Subsequently, Section~4.2 of that paper made allowance for the idea that, while estimates of different quantities might be correlated, the estimates {\em of the corresponding underlying variances} might remain independent, and the rest of the paper was presented from that more general point. However, arguably, dependence between such ${\pmb x}_i$ estimates arises in practice from simultaneous sampling, and the appropriate estimates of variance will also be dependent. Therefore, (\ref{300}) is of considerable importance in its own right,\footnote{For example, this result is relevant when estimating the parameters of a straight line in `ordinary least-squares regression'. The uncertainties of the intercept and slope are based on the same sum of squares of the residuals. Similarly, the result is applicable when measuring a function of a complex quantity from the results of $n$ simultaneous measurements of its real and complex parts, and where the variances are estimated using the spread of the results.} and here we extend that result to the multidimensional case.

\vskip 3mm
\noindent
{\em Analogous result for the multivariate case}\\[2mm]
\noindent
To develop the analogous result for the multidimensional case we will use a relationship that can act as the definition of the distribution of Hotelling's $T^2$ \cite{classical}.
\begin{quote}
If ${\pmb z}\leftarrow {\rm N}_p({\bf 0},{\boldsymbol\Sigma})$ and independently
${\bf v}\leftarrow \frac{1}{\nu}{\rm W}_p(\nu,{\boldsymbol\Sigma})$ then the scalar ${\pmb
z}^\prime{\bf v}^{-1}{\pmb z }$ is drawn from the distribution of Hotelling's
$T^2$ with $\nu$ degrees of freedom and with second parameter $p$, i.e. ${\pmb
z}^\prime{\bf v}^{-1}{\pmb z }\leftarrow {\rm T}^2_{\nu, p}$.\\[-8mm]
\begin{equation}\label{400}\end{equation}
\end{quote}

Suppose that a significant component of error in a multivariate measurement is given by a known linear combination of $h$ individual errors, $\sum_{j=1}^h {\bf c}_j ({\pmb x}_j - {\pmb X}_j)$, and suppose the ${\pmb x}_j$ estimates have been formed from $n_{}$ independent sets of multivariate observations. Then, for the mathematics, it is helpful to write
\[
\sum_{j=1}^h {\bf c}_j ({\pmb x}_j - {\pmb X}_j)={\bf c}_{} ({\pmb x}_{} - {\pmb X}_{}),
\]
where ${\pmb X}_{}$ is the column vector of length $h{q}$ formed by concatenating the vectors ${\pmb X}_{1},\ldots,{\pmb X}_h$, ${\pmb x}_{}$
is formed by concatenating ${\pmb x}_{1},\ldots,{\pmb x}_h$, and ${\bf c}_{}$ is equal to $[{\bf c}_1,\ldots,{\bf c}_h]$.
Consequently, we can write ${\pmb x}_{}-{\pmb X}_{} \leftarrow {\rm N}_{hr}\left({\bf 0},{\bf \Sigma}_{}/n_{}\right)$
for some covariance matrix ${\boldsymbol \Sigma}_{}$. Thus,
\begin{equation}
{\bf c}_{} ({\pmb x}_{} - {\pmb X}_{}) \leftarrow {\rm N}_p \left({\bf 0},{\bf c}_{}{\bf \Sigma}_{}{\bf c}_{}^\prime/n_{} \right)
\label{numer}
\end{equation}
and the $hr\times h{q}$ sample covariance matrix ${\bf s}_{}$ obeys
\[
(n_{}-1){\bf s}_{}\leftarrow {\rm W}_{hr}(n_{}-1,{\boldsymbol\Sigma}_{}).
\]
So, from (\ref{100}),
\begin{equation}
(n_{}-1){\bf c}_{}{\bf s}_{}{\bf c}_{}^\prime \leftarrow {\rm W}_p(n_{}-1,{\bf c}_{}{\boldsymbol\Sigma}_{}{\bf c}_{}^\prime).
\label{denom}
\end{equation}
The vector on the left of (\ref{numer}) is drawn independently of the matrix on the left of (\ref{denom}) so, from (\ref{400}),
\[
\{{\bf c}_{} ({\pmb x}_{} - {\pmb X}_{})\}^\prime\{{\bf c}_{}{\bf s}_{}{\bf c}_{}^\prime\}^{-1}\{{\bf c}_{} ({\pmb x}_{} - {\pmb X}_{})\}/(n_{}-1) \leftarrow {\rm T}^2_{n_{}-1,p}.
\]
This relationship is exact. It implies that the appropriate number of degrees of freedom for an estimate of the covariance matrix of a fixed linear combination of vector-valued errors obtained from $n_{}$ simultaneous measurements is $n_{}-1$. This result means that (\ref{300}) is a valid statement in multivariate problems also.

\vskip 3mm
\noindent
{\em Generalization}\\[2mm]
\noindent
Let us now see how this result applies with (\ref{linapprox}) where all of the $m$ measurement results are obtained by sampling from normal distributions, with some being obtained in the way described and others being obtained independently.

Suppose $h$ of the ${\pmb X}_i$ quantities are measured together on $n_{m+1}$ independent occasions, and suppose the labelling is such that these are ${\pmb X}_{1},\ldots,{\pmb X}_h$. So  $n_{m+1}=n_1=\cdots=n_h$. Let ${\pmb X}_{m+1}$ be the vector of length $h{q}$ formed by concatenating the vectors ${\pmb X}_{1},\ldots,{\pmb X}_h$ and let ${\pmb x}_{m+1}$ be the vector formed by concatenating the vectors ${\pmb x}_{1},\ldots,{\pmb x}_h$. We form the $p\times h{q}$ matrix ${\bf c}_{m+1}=[{\bf c}_{1},\ldots,{\bf c}_h]$ and, from (\ref{linapprox}), write
\[
{\pmb y}-{\pmb Y}\approx \sum_{i=h+1}^{m} {\bf c}_i ({\pmb x}_i-{\pmb X}_i)+ {\bf c}_{m+1} ({\pmb x}_{m+1}-{\pmb X}_{m+1}).
\]
The estimate of  ${\pmb X}_{m+1}$ obtained on each occasion arises from a multivariate normal distribution with mean vector ${\pmb X}_{m+1}$ and some unknown $hr\times h{q}$ covariance matrix  ${\bf \Sigma}_{m+1}$. So the sample mean ${\pmb x}_{m+1}$ obeys (\ref{numer}) and the corresponding sample covariance matrix ${\bf s}_{m+1}$ obeys (\ref{denom}), (with the subscript $m+1$ inserted where appropriate). Define
\[
{\bf w}_{m+1} = {\bf c}_{m+1}{\bf v}_{m+1}{\bf c}_{m+1}^\prime.
\]
Then the realised confidence region is given by (\ref{basicregion}) but now with ${\bf v}_{\pmb y} = \sum_{i=h+1}^{m+1}  {\bf w}_i$ and with $\nu_{\rm eff}$ calculated from (\ref{nueff14}), where summation in (\ref{nueff14}) is now from $i=h+1$ to $i=m+1$.

Similarly, for the bivariate case of $p=2$, $\nu_{\rm eff}$ is calculated from (\ref{ADFadf}) with the limits of summation over $i$ now being $h+1$ and $m+1$.

\subsection{Example 2}
\label{example2sec}

Consider again the measurement problem described in Example~1, where the complex reflection coefficient $\Gamma$ is measured using (\ref{Gamma}). We no longer assume that the denominator of that equation can be approximated by $1+0{\rm i}$, so we progress from an analysis with $m=2$ to an analysis in which $m=5$, where $\Gamma$ is estimated by
\begin{equation}
\hat{\Gamma} = \frac{\hat{\Gamma}^\prime - \hat{S}_{11}}{\hat{S}_{12}\hat{S}_{21} + \hat{S}_{22}(\hat{\Gamma}^\prime - \hat{S}_{11})}.
\label{hatGammaEx2}
\end{equation}

Standard instrumentation like that generally found in RF metrology laboratories can repeatedly make simultaneous measurements of the four $S$-parameters. So, in one step of a measurement procedure to estimate $\Gamma$, a number of measured values of $S_{11}$, $S_{12}$, $S_{21}$ and $S_{22}$ will be obtained. Here ${\pmb X}_1={S}_{11}$, ${\pmb X}_2={S}_{12}$, ${\pmb X}_3={S}_{21}$, ${\pmb X}_4={S}_{22}$ and $h=4$.
In this case, the data on these vectors are obtained simultaneously and dependence between the errors is envisaged, so these vectors are concatenated into the vector
\begin{equation}
{\pmb X}_6=\left({S}_{11. {\rm re}},{S}_{11. {\rm im}},{S}_{12. {\rm re}},{S}_{12. {\rm im}},
{S}_{21. {\rm re}},{S}_{21. {\rm im}},{S}_{22. {\rm re}},{S}_{22. {\rm im}}
\right)^\prime,
\label{X6}
\end{equation}
where the real and imaginary components of the complex numbers are indicated. From this data, the sample mean vector ${\pmb x}_6$ and the corresponding $8\times 8$ covariance matrix can be calculated.  The sample mean vector ${\pmb x}_6$ contains the estimates of the $S$-parameters, $\hat{S}_{11}$, $\hat{S}_{12}$, $\hat{S}_{21}$ and $\hat{S}_{22}$. Independently, in another step of the procedure, a number of measured values for $\Gamma^\prime$, which is ${\pmb X}_5$, will be obtained. From this data, the sample mean ${\pmb x}_5 =\hat{\Gamma}^\prime$ and the corresponding $2\times 2$ covariance matrix will be calculated.

%%%%TABLE 1
\begin{table}[h]
\caption{Results obtained in 7 simultaneous measurements of the $S$-parameters}
\begin{center}
\begin{tabular}{ccccccc}
$S_{11}$&&$S_{12}$&&$S_{21}$&&$S_{22}$\\
\hline
$-0.072+0.066\,\mathrm{j}$&&$0.112+0.903\,\mathrm{j}$&&$ 0.106+0.900\,\mathrm{j}$&&$ 0.044+0.106\,\mathrm{j}$\\
$-0.075+0.073\,\mathrm{j}$&&$ 0.115+0.891\,\mathrm{j}$&&$ 0.098+0.900\,\mathrm{j}$&&$ 0.029+0.091\,\mathrm{j}$\\
$-0.087+0.069\,\mathrm{j}$&&$ 0.094+0.888\,\mathrm{j}$&&$ 0.085+0.893\,\mathrm{j}$&&$ 0.019+0.096\,\mathrm{j}$\\
$-0.087+0.092\,\mathrm{j}$&&$ 0.098+0.898\,\mathrm{j}$&&$ 0.106+0.886\,\mathrm{j}$&&$ 0.036+0.099\,\mathrm{j}$\\
$-0.064+0.071\,\mathrm{j}$&&$ 0.092+0.899\,\mathrm{j}$&&$ 0.114+0.907\,\mathrm{j}$&&$ 0.026+0.102\,\mathrm{j}$\\
$-0.079+0.067\,\mathrm{j}$&&$ 0.072+0.892\,\mathrm{j}$&&$ 0.089+0.882\,\mathrm{j}$&&$ 0.026+0.088\,\mathrm{j}$\\
$-0.069+0.072\,\mathrm{j}$&&$ 0.102+0.906\,\mathrm{j}$&&$ 0.108+0.893\,\mathrm{j}$&&$ 0.022+0.096\,\mathrm{j}$\\
\hline
\end{tabular}
\end{center}
\end{table}

Suppose that the first step of the procedure is carried out using 7 measurements and that the results are those indicated in Table~1. So $n_6=7$, the
estimates of the $S$-parameters are $\hat{S}_{11}= -0.0761+0.0729\,\mathrm{j}$, $\hat{S}_{12}= 0.0979+0.8967\,\mathrm{j}$,  $\hat{S}_{21}= 0.1009+0.8944\,\mathrm{j}$ and $\hat{S}_{22}= 0.0289+0.0969\,\mathrm{j}$, and the sample mean vector is
\[
{\pmb x}_6=  (-0.0761,\,  0.0729,\,  0.0979,\,  0.8967,\,  0.1009,\,  0.8944,\,  0.0289,\,  0.0969)^\prime.
\]
Also, suppose that the second step of the procedure is carried out using 5 measurements, and that the results are $-0.220+0.008\,\mathrm{j},\,-0.217-0.010\,\mathrm{j},\,-0.222-0.003\mathrm{j},\,-0.207-0.018\mathrm{j}$ and $-0.219-0.009\mathrm{j}$. So $n_5=5$, $\hat{\Gamma}^\prime= -0.2170-0.0064\mathrm{j}$ and ${\pmb x}_5=(-0.2170, -0.0064)^\prime$.  From (\ref{hatGammaEx2}), the measured value of $\Gamma$ is
\[
\hat{\Gamma} = 0.1516 + 0.1317 \,\mathrm{j}.
\]

The estimate of the parent covariance matrix of ${\pmb x}_5$ is calculated to be
\[
{\bf v}_5 =
\left[
\begin{array}{rr}
0.6900 & -0.8550\\
-0.8550 &1.8660\\
\end{array}
\right]\times 10^{-5}
\]
and the estimate of the parent covariance matrix of ${\pmb x}_5$ is
\[
{\bf v}_6 =
\left[
\begin{array}{rrrrrrrr}
1.0973 & -0.4908 & $\;$0.3592 & $\;$0.4946 & $\;$0.9020 & $\;$0.7486 & $\;$0.0401 & $\;$0.2354\\
-0.4908 & $\;$1.1116 & 0.1949 & 0.0707 & 0.3878 & -0.3395 & 0.2354 & 0.0568\\
0.3592 & 0.1949 & 2.9259 & 0.4088 & 0.8211 & 1.0010 & 0.7568 & 0.5425\\
0.4946 & 0.0707 & 0.4088 & 0.6272 & 0.8231 & 0.1878 & 0.3160 & 0.3493\\
0.9020 & 0.3878 & 0.8211 & 0.8231 & 1.6116 & 0.7010 & 0.5187 & 0.6068\\
0.7486 & -0.3395 & 1.0010 & 0.1878 & 0.7010 & 1.0707 & 0.1153 & 0.4224\\
0.0401 & 0.2354 & 0.7568 & 0.3160 & 0.5187 & 0.1153 & 1.0497 & 0.4235\\
0.2354 & 0.0568 & 0.5425 & 0.3493 & 0.6068 & 0.4224 & 0.4235 & 0.5449\\
\end{array}
\right]\times 10^{-5}.
\]
Also, from (\ref{cimatrixcomplex}) we find that
\[
{\bf c}_5=
\left[
\begin{array}{rr}
-1.2316 & 0.2296\\
-0.2296 &-1.2316
\end{array}
\right]
\]
and from (\ref{cimatrix1})--(\ref{cimatrix2}), (or from repeated application of (\ref{cimatrixcomplex})), we obtain
\[
{\bf c}_6 =
\left[
\begin{array}{rrrrrrrr}
1.2316 &-0.2296 &-0.1619 &-0.1555 &-0.1628& -0.1552& -0.0056 & 0.0399\\
0.2296  &1.2316 & 0.1555 &-0.1619&  0.1552 &-0.1628 &-0.0399& -0.0056
\end{array}
\right].
\]
Then (\ref{wcuc}) is applied to give
\[
{\bf w}_5=
\left[
\begin{array}{rr}
1.629 & -1.584\\
-1.584&  2.383\\
\end{array}
\right]\times 10^{-5}
\hspace{1cm}{\rm and}\hspace{1cm}
{\bf w}_6=
\left[
\begin{array}{rr}
1.406 & -0.999\\
-0.999 & 1.816\\
\end{array}
\right]\times 10^{-5},
\]
and from (\ref{vy12}) we obtain
\[
{\bf v}_y=
\left[
\begin{array}{rr}
3.035 & -2.583\\
-2.583&  4.199\\
\end{array}
\right]\times 10^{-5}.
\]
Equation (\ref{ADFadf}) gives $\nu_{\rm eff}=9.01$ so the critical value on the right-hand side of (\ref{basicregion}) is 10.03. The corresponding elliptical 95\% region of measurement uncertainty for the measurand $\Gamma$ is the small ellipse shown in Figure~\ref{ellipses}. The measurement result $\hat{\Gamma}$ is marked at the centre of this region.

\vskip 3mm
\noindent
{\em Comparison with the method of {Supplement 2}}\\[2mm]
\noindent
We saw in Example~1 that, when a fixed $q$-dimensional quantity is estimated by sampling, {\em Supplement 2} recommends subsequent description of that quantity using the distribution with joint probability density function (\ref{posterior}), which is only defined
for $n_i>q$. Example~2 involves 7 simultaneous measurements of the 8 components of the $S$-parameters, so $n_i=7$ and we are measuring a quantity with $q=8$. Therefore the probability density function (\ref{posterior}) cannot be attributed to the vector of $S$-parameters and the method recommended in {\em Supplement 2} to the {\em Guide} is inapplicable.

To employ the method of {\em Supplement 2} there must be at least two more simultaneous measurements of the $S$-parameters. By the same principles that govern the size of the coverage region in Section~\ref{example1sec}, where $X_i$ is attributed the covariance matrix (\ref{bigcov}), the typical coverage region for $\Gamma$ would then be considerably larger than the corresponding region obtained by our method.

\subsection{Performance assessment}

In this section we study the performance of the method in the situation underlying Example~2 for reasonable settings of the parameters that govern the measurement. The measurement process can be simulated when we have chosen values for the four $S$-parameters and for $\Gamma$ and when we have defined the process in which the estimates of the $S$-parameters and $\Gamma^\prime$ are obtained. Application of the Monte Carlo principle then enables aspects of performance to be examined; the two most relevant being (i) the frequency with which the process leads to a region containing $\Gamma$, i.e. the `success rate',  and (ii) some measure of the typical size of this region.

The vector ${\pmb X}_6$ in (\ref{X6}) contains the real and imaginary parts of the $S$-parameters. In any simulated measurement, the data on these parameters are $n_6$ individual estimates of the elements of ${\pmb X}_6$. The corresponding estimates are the elements of the associated sample mean vector ${\pmb x}_6$ and the relevant uncertainty information is contained in a corresponding $8\times 8$ sample covariance matrix. The data on the $S$-parameters were drawn from the multivariate normal distribution with mean vector ${\pmb X}_6$, variances $\sigma^2_{\rm re}$ and $\sigma^2_{\rm im}$ for real and imaginary components, and correlation matrix
\begin{equation}
\left[
\begin{array}{cc|cc|cc|cc}
1&\rho&\kappa&\kappa&\kappa&\kappa&\kappa&\kappa\\
\rho&1&\kappa&\kappa&\kappa&\kappa&\kappa&\kappa\\
\hline
\kappa&\kappa&1&\rho&\kappa&\kappa&\kappa&\kappa\\
\kappa&\kappa&\rho&1&\kappa&\kappa&\kappa&\kappa\\
\hline
\kappa&\kappa&\kappa&\kappa&1&\rho&\kappa&\kappa\\
\kappa&\kappa&\kappa&\kappa&\rho&1&\kappa&\kappa\\
\hline
\kappa&\kappa&\kappa&\kappa&\kappa&\kappa&1&\rho\\
\kappa&\kappa&\kappa&\kappa&\kappa&\kappa&\rho&1\\
\end{array}
\right],
\label{firstmatrix}
\end{equation}
in which $\rho$ is the correlation coefficient between the estimators of the real and imaginary components of individual $S$-parameters, and $\kappa$ is the correlation coefficient between the estimators of each pair of components of different $S$-parameters.

Similarly, the vector ${\pmb X}_5 \equiv \left({\Gamma}^\prime_{\rm re},{\Gamma}^\prime_{\rm im}\right)^\prime$ contains the real and imaginary parts of $\Gamma^\prime$. The data on $\Gamma^\prime$ are $n_5$ individual estimates of the elements of ${\pmb X}_{\rm 5}$. The corresponding measurement results are the elements of the sample mean vector ${\pmb x}_{\rm 5}\equiv \left(\hat{\Gamma}^\prime_{\rm re},\hat{\Gamma}^\prime_{\rm im} \right)^\prime$ and the uncertainty information is contained in the corresponding $2\times 2$ sample covariance matrix. The data were drawn from the multivariate normal distribution with mean vector ${\pmb X}_5$ and covariance matrix
\[
\left[
\begin{array}{cc}
\sigma^2_{\rm re}&\rho\,\sigma_{\rm re}\,\sigma_{\rm im}\\
\rho\,\sigma_{\rm re}\,\sigma_{\rm im}&\sigma^2_{\rm im}\\
\end{array}
\right],
\]
with $\rho$ having the same value as in (\ref{firstmatrix}).

The model of the measurement system is determined once the amplitudes and phases of the $S$-parameters are specified. To limit the numerical experiment to a manageable size we fixed the magnitudes of the $S$-parameters to be  $|S_{11}|=|S_{22}|=0.1$ and $|S_{12}|=|S_{21}|=0.9$. Also we set $n_6=n_5=n$ and $\sigma_{\rm re}=0.01$, and then used a variable ratio ${\tau} \equiv \sigma_{\rm im}/\sigma_{\rm re}$ to define the value of $\sigma_{\rm im}$. For each combination of $|\Gamma| \in \{0.2,0.5,0.8\}$, $n \in \{4,8,16\}$, ${\tau} \in \{0.5,1.0,2.0\}$, $\rho \in \{-0.9,-0.6,-0.3,0,0.3,0.6,0.9\}$ and $\kappa \in \{0,\;\rho/2\;{\rm if}\;\rho>0,\;\rho\;{\rm if}\;\rho>0\}$, fifty sets of random phases for $S_{11}$, $S_{22}$ and $S_{12}$ were generated, and in each case the phase of $S_{21}$ was set equal to the phase of $S_{12}$. (So in all the simulations $S_{12}=S_{21}$.) In this way, we were defining fifty ${\pmb X}_6$ vectors to explore the space of possible measurement situations for each setting of the other parameters. For each of the $3\times 3 \times 3 \times (4 + 3\times 3) \times 50 = 17550$ situations so defined, $10^4$ $n$-fold measurements were simulated in order to compare the performance of the method proposed in this paper with the performances of two other methods of analysis. Each measurement involved selecting a random phase for $\Gamma$. This means that the performance measures obtained relate to the behaviour of the system when measuring many different measurands with the same magnitude $|\Gamma|$ but with different phases, rather than in repeated measurement of the same measurand, which is a useful but hypothetical concept.

The three methods of analysis studied are as follows:\\

\noindent
{\bf Method 1 (the method proposed in this paper):} As has been described above, the proposed method involves acknowledging that the data on each of the four S-parameters are obtained simultaneously. In this case $h=4$, $m=5$ and the method is applicable with $i$ being summed from $i=5$ to $i=6$. (This method has been proposed for the situation where no justifiable assumptions could be made about $\rho$ and $\kappa$.)\\

\noindent
{\bf Method 2:}
The second method corresponds to using the analysis given by Willink and Hall \cite{classical}. Here dependence between the estimators of the real and imaginary components of a complex quantity is taken into account, but it is assumed that there is no dependence between the estimators of different complex quantities, as if the four $S$-parameters were measured independently. In this case the method is applicable with $i$ being summed from $i=1$ to $i=5$. (So this method was proposed for the situation where it could justifiably be assumed that $\kappa=0$.)\\

\noindent
{\bf Method 3:} The third method involves forming separate 97.5\% intervals of uncertainty for the real and imaginary components of $\Gamma$
by applying the method of the {\em Guide} with the W-S formula to the ten quantities forming the elements of ${\pmb X}_6$ and ${\pmb X}_5$. In this process, the off-diagonal elements of the sample covariance matrices are disregarded. By the Bonferroni inequality, see e.g.\ \cite{MUAP}, the two intervals form a valid rectangular 95\% uncertainty region for $\Gamma$; the word `valid' meaning that the success rate of the procedure is at least 95\%. (In effect, this method involves the assumption that $\rho=0$ and $\kappa=0$.)\\

For $k=1,2,3$, the performance characteristics studied for Method $k$ are (i) the proportion of occasions on which the uncertainty region calculated on the complex plane contained $\Gamma$, which is the success rate $p_k$, and (ii) the typical area of the region, as represented by the ratio of the average area of this region $\bar{A}_k$ to the average area in the region calculated by one of the other methods. (With there being $10^4$ trials for each setting of the parameters, a recorded success rate of $p_k$ close to 0.95 can be regarded as being accurate to $\pm 2\sqrt{0.95\times 0.05/10^4} = \pm 0.004$. The area of the ellipse defined by (\ref{basicregion}) when $p=2$ is $A=\pi c \sqrt{{\rm det}\,{\bf v}_{\pmb y}}$, where $c$ is the critical value on the right-hand side of (\ref{basicregion}), see e.g.\ \cite[eq. 4.7]{JohnsonWichern}.)

Apart from the issue of there being edge effects with reflection coefficients with magnitudes near zero or one, the success rates and area ratios will not depend on $\sigma_{\rm re}$, so the results obtained with the single value of $\sigma_{\rm re}=0.01$ used in the simulations are applicable with other reasonable values of $\sigma_{\rm re}$ also.

Table~2 describes the performances of the methods when $|\Gamma|=0.2$, and presents the results in three parts for $n=4$, $n=8$ and $n=16$, in order from top to bottom. Each line gives results obtained by considering the scenarios that involved the values of $\rho$ and $\kappa$ given in the first two columns. Varying $\tau$ was found to make little difference, so the results for the three values of $\tau$ were grouped together to obtain a set of 150 scenarios for each combination of $\rho$ and $\kappa$. The performance figures listed in each line are the minimum and maximum values of $p_1$, $p_2$ and $p_3$ and the minimum and maximum values of the ratios $\bar{A}_1/\bar{A}_2$, $\bar{A}_1/\bar{A}_3$ and $\bar{A}_2/\bar{A}_3$, where these maxima and minima are taken over these 150 scenarios. (For example, the first row in Table~2 shows that, when $|\Gamma|=0.2, n=4, \rho=0$ and $\kappa=0$, for each scenario the success rate observed with Method~1 was at least 96.3\%, for each scenario the success rate observed with Method~3 was at least 95.1\%, and for each scenario the mean area of the region generated using Method~1 was at least 1.23 times the mean area of the region generated using Method~3.) Table~3 gives the corresponding results for $|\Gamma|=0.8$. As might be expected, the figures for $|\Gamma|=0.5$ lay between the figures for $|\Gamma|=0.2$ and $|\Gamma|=0.8$ in the vast majority of cases, and those that did not could be associated with random variation. So, with the exception of the subset of figures given in Table~4, the results for $\Gamma=0.5$ are not shown.

Our main interest resides in the minimum observed success rates, which should be approximately 95\% or more if we are to claim that the method generates intervals of measurement uncertainty in which there can be 95\% trust.  The figures that might be expected to drop below 95\% are indicated in bold type. These correspond to situations for which the assumptions involved in the methods are not satisfied. So, in Tables~2, 3 and 4, there are bold figures of minimum success rate for Method 2 whenever $\kappa\ne 0$ and for Method 3 whenever $\rho\ne 0$.

Over all the scenarios studied, Method~1 had a minimum success rate of 94.1\%, (which was a figure observed with $|\Gamma|=0.5$ and so is not seen in Tables~2 and 3). Thus, in accordance with the theory given in this paper, the method may be said to generate in practice `95\% uncertainty intervals'. However, the price to pay for this robustness to the presence of correlations between the error processes is an enlargement of the uncertainty region. It is clear from the ratios of mean areas that Method~1 should only be applied when Methods 2 and 3 are deemed to be insufficiently reliable.

Tables~2 and 3 show that when there was no correlation between any of the estimates, i.e.\ when $\rho=\kappa=0$, the success rate of Method 3 was approximately 95\% or more and the regions obtained were on average considerably smaller than those obtained using Methods~1 and 2.  One reasonable conclusion is that when the error in each individual estimate is thought to have been generated independently or nearly independently, Method 3 is to be preferred. The tables also show that the success rate with Method 3 dropped considerably below 0.95 when there were correlated errors, so that Method 3 must be deemed unreliable if significant correlation is suspected.

When correlation existed between the estimates of real and imaginary components of the same complex quantity but not between the estimates of the components of different complex quantities, i.e. when $\rho\ne 0$ but $\kappa=0$, Method~2 had a success rate approximately 95\% or more and generated regions smaller than those obtained using Method~1. This is to be expected from theory. These results give further support to the use of Method 2 for such problems.

Furthermore, Method~2 showed some robustness to the presence of correlation between the estimates of the components of different quantities, i.e. to non-zero $\kappa$. Table~2 shows that its success rate was entirely adequate when $|\Gamma|=0.2$ with $n=4$ and in most cases with $n=8$ and $n=16$, while it continued to give smaller regions than Method~1. Table~3 shows this was not the case for $|\Gamma|=0.8$. Table~4 shows the corresponding minimum success rates when $|\Gamma|=0.5$. Most of the minimum success rates with $n=4$ are close to 95\%. So Method~2 performed well in some situations for which it was not designed, notably with small $n$.

\addtocounter{table}{2}
%%%%TABLE 4
\begin{table}[h]
\caption{Results for $|\Gamma|=0.5$. Minimum success rates with Method 2 for the scenarios with $\kappa\ne 0$.}
{\small
\begin{center}
\begin{tabular}{rlcccc}
$\rho$&$\kappa$&&$n=4$&$n=8$&$n=16$\\
\hline
0.3 & 0.15 && {\bf 96.1} & {\bf 94.7} &{\bf 94.1 }\\
0.3 & 0.3 && {\bf 95.6} & {\bf 93.9} &{\bf 93.5 }\\
0.6 & 0.3 && {\bf 95.7} & {\bf 94.1} &{\bf 93.5 }\\
0.6 & 0.6 && {\bf 94.9} & {\bf 92.8} &{\bf 91.9 }\\
0.9 & 0.45 && {\bf 95.7} & {\bf 93.3} &{\bf 92.5 }\\
0.9 & 0.9 && {\bf 93.3} & {\bf 90.0} &{\bf 89.4 }\\
\hline
\end{tabular}
\end{center}
}
\end{table}

\section{The general formula}

The main results of this paper can be brought together into a single form and generalized to the situation where there are several groups of simultaneously measured quantities. The general result is as follows:
\begin{quote}
Consider measuring a quantity  ${\pmb Y}=f({\pmb X}_1,\ldots,{\pmb X}_m)$ where ${\pmb Y}$ is a column vector of length $p$ and each ${\pmb X}_i$ is a column vector of length ${q}\ge p$. Suppose that the result is ${\pmb y}=f({\pmb x}_1,\ldots,{\pmb x}_m)$ where ${\pmb x}_i$ is the average of $n_i$ measurements having sample covariance matrix ${\bf s}_i$. Also suppose that the error ${\pmb y}-{\pmb Y}$ can be adequately approximated by the first-order expression ${\pmb y}- {\pmb Y}\approx \sum_{i=1}^m {\bf c}_i ({\pmb x}_i-{\pmb X}_i)$. Set
\[
{\bf v}_i \equiv \frac{{\bf s}_i}{n_i}
\]
and
\[
{\bf w}_i \equiv\left(w_{i\cdot jk}\right) = {\bf c}_i{\bf v}_i{\bf c}_i^\prime,
\]
and let the figure $F_{{p},\nu_{\rm eff}+1-{p},0.95}$ denote the 0.95 quantile of the $F$-distribution with parameters ${p}$ and $\nu_{\rm eff}+1-{p}$ in the numerator and denominator. Then an approximately 95\% uncertainty region for ${\pmb Y}$ is the interior of the hyper-ellipsoid defined by the locus of vectors ${\pmb Y}^\ast$ satisfying
\begin{equation}
({\pmb{y}}-{\pmb Y}^\ast)^\prime {\textstyle\left(\sum_i {\bf w}_i\right)}^{-1} ({\pmb{y}}-{\pmb Y}^\ast)
=\frac{\nu_{\rm eff}{p}}{\nu_{\rm eff}+1-{p}}F_{{p},\nu_{\rm eff}+1-{p},0.95}
\label{finalregion}
\end{equation}
where
\begin{equation}
\nu_{\rm eff}=\frac{\displaystyle \sum_{j=1}^p\sum_{k=j}^p \left\{\sum_{i} w_{i\cdot jj}\sum_{i} w_{i\cdot kk}+\left(\sum_{i} w_{i\cdot
jk}\right)^2\right\}}{\displaystyle \sum_{j=1}^p\sum_{k=j}^p\left\{\sum_{i}\left(w_{i\cdot jj}w_{i\cdot kk}+w_{i\cdot jk}^2\right)/(n_i-1)\right\}}
\label{nueff19}
\end{equation}
and where the range of summation over $i$ is described in points (i)-(iii) below. In the important case of $p=2$, (\ref{nueff19}) simplifies to
\begin{equation}
\nu_{\rm eff}=\frac{
2\left(\sum_{i} w_{i\cdot 1{}1}\right)^2+
\sum_{i} w_{i\cdot 1{}1}\sum_{i} w_{i\cdot 2{}2}+\left(\sum_{i} w_{i\cdot 1{}2}\right)^2+
2\left(\sum_{i} w_{i\cdot 2{}2}\right)^2}{
\sum_{i} \left(2 w_{i\cdot 1{}1}^2 + w_{i\cdot 1{}1}w_{i\cdot 2{}2}+w_{i\cdot 1{}2}^2+2 w_{i\cdot 2{}2}^2\right)/(n_i-1).
}
\label{eq24}
\end{equation}

(i) In the basic situation, each set of $n_i$ measurement results is obtained in separate experiments. In that case the summation over $i$ in (\ref{finalregion})-(\ref{eq24}) runs from $i=1$ to $i=m$.

(ii) However, we can also envisage the situation where the first $h$ estimates are obtained from $n_{m+1}$ simultaneous measurements of ${\pmb X}_{1},\ldots,{\pmb X}_{h}$, (so $n_{m+1}=n_1=\ldots=n_h$). In this situation, we set
\[
{\bf w}_{m+1} = {\bf c}_{m+1}{\bf v}_{m+1}{\bf c}_{m+1}^\prime
\]
where ${\bf c}_{m+1}=\left[{\bf c}_{1},\cdots,{\bf c}_h\right]$, ${\bf v}_{m+1}={\bf s}_{m+1}/n_{m+1}$ and ${\bf s}_{m+1}$ is the $h{q}\times h{q}$ sample covariance matrix for the $n_{m+1}$ individual estimates of the column vector $\left({\pmb X}_{1}^\prime,\cdots,{\pmb X}_{h}^\prime\right)^\prime$. Then the summation over $i$ in (\ref{finalregion})-(\ref{eq24}) runs from $i=h+1$ to $i=m+1$.

(iii) We can also envisage the more general situation where the first $h_1$ estimates are obtained from $n_{m+1}$ simultaneous measurements of ${\pmb X}_{1},\ldots,{\pmb X}_{h_1}$, the next $h_2$ estimates are obtained from $n_{m+2}$ simultaneous measurements of ${\pmb X}_{h_1+1},\ldots,{\pmb X}_{h_1+h_2}$, and so on. Suppose there are $k$ such groups of estimates. For $i=1,\ldots,k$ we set
\[
{\bf w}_{m+i} = {\bf c}_{m+i}{\bf v}_{m+i}{\bf c}_{m+i}^\prime
\]
where ${\bf c}_{m+i}$ is the $p\times h_i{q}$ matrix given by
\[
{\bf c}_{m+i}=\left[{\bf c}_{1+\sum_{j=1}^{i-1} h_j},\cdots,{\bf c}_{\sum_{j=1}^{i} h_j}\right],
\]
${\bf v}_{m+i}={\bf s}_{m+i}/n_{m+i}$ and ${\bf s}_{m+i}$ is the $h_i{q}\times h_i{q}$ sample covariance matrix for the $n_{m+i}$ individual estimates of the column vector
\[
\left({\pmb X}_{1+\sum_{j=1}^{i-1} h_j}^\prime,\cdots,{\pmb X}_{\sum_{j=1}^{i} h_j}^\prime\right)^\prime.
\]
Then the summation over $i$ in (\ref{finalregion})-(\ref{eq24}) runs from $i=\sum_{j=1}^k h_j+1$ to $i=m+k$.
\end{quote}

When the measurand and the input quantities are one-dimensional, i.e. when $p={q}=1$, the general procedure above reduces to existing procedures. In that situation, the measurand is a scalar $Y$, the input quantities are scalars $X_1,\ldots,X_m$, and the measurement estimate is a scalar $y$. The matrices ${\bf w}_i$ and ${\bf c}_i$ are scalars, say $w_i$ and $c_i$, while ${\bf v}_i$ is the scalar $u_i^2=s_i^2/n_i$.  Taking the square root of each side of (\ref{finalregion}) then gives the expression
\[
\pm({y}-{Y}^\ast)\left(\sum_i {w}_i\right)^{-1/2} = t_{\nu_{\rm eff},0.95}
\]
where $t_{\nu_{\rm eff},0.95}$ is the 0.95 quantile of the $t$-distribution with $\nu_{\rm eff}$ degrees of freedom. The 95\% region of measurement uncertainty is then the interval with limits
\begin{equation}
y \pm t_{\nu_{\rm eff},0.95} \times \sqrt{\sum_i c_i^2{u}^2_i}.
\label{finalinterval}
\end{equation}
Also equation (\ref{nueff19}) reduces to
\begin{equation}
\nu_{\rm eff}=\frac{\displaystyle \left(\sum_i c_i^2{u}^2_i\right)^2}{\displaystyle \sum_i \left(c_i^4{u}_i^4\right)/(n_i-1)},
\label{nueff20}
\end{equation}
which is in the form of the W-S formula \cite[appendix G.4]{GUM} but with summation over $i$ extending over an unspecified range.

Suppose each set of $n_i$ measurement results is obtained in separate experiments, as in (i) above. Then the summation over $i$ in (\ref{finalinterval}) and (\ref{nueff20}) runs from $i=1$ to $i=m$, and the procedure is simply the basic procedure of the {\em Guide} for independent one-dimensional components. Alternatively, suppose that the first $h_1$ results are obtained from $n_{m+1}$ simultaneous measurements of ${X}_{1},\ldots,{X}_{h_1}$, the next $h_2$ results are obtained from $n_{m+2}$ simultaneous measurements of ${X}_{h_1+1},\ldots,{X}_{h_1+h_2}$, and so on, as in (iii) above. Suppose there are $k$ such groups of results. For $i=1,\ldots,k$ we set
\[
{w}_{m+i} = {\bf c}_{m+i}{\bf v}_{m+i}{\pmb c}_{m+i}^\prime
\]
where ${\bf c}_{m+i}$ is the $1\times h_i$ matrix (i.e. a row vector) given by
\[
{\bf c}_{m+i}=\left[{c}_{1+\sum_{j=1}^{i-1} h_j},\cdots,{c}_{\sum_{j=1}^{i} h_j}\right],
\]
${\bf v}_{m+i}={\bf s}_{m+i}/n_{m+i}$ and ${\bf s}_{m+i}$ is the $h_i\times h_i$ sample covariance matrix for the $n_{m+i}$ individual estimates of the column vector
\[
\left({X}_{1+\sum_{j=1}^{i-1} h_j}^\prime,\cdots,{X}_{\sum_{j=1}^{i} h_j}^\prime\right)^\prime.
\]
Then the summation over $i$ in (\ref{finalinterval}) and (\ref{nueff20}) runs from $i=\sum_{j=1}^k h_j+1$ to $i=m+k$. (This is in accordance with Result~1 of Willink \cite{generalize}, which we have expressed as (\ref{300}).)

The analysis above is also applicable when the measurand is univariate but the input quantities are multivariate, i.e. when $p=1$ but ${q}>1$. In that case $w_i  = {\bf c}_i{\bf v}_i{\bf c}_i^\prime$ and the 95\% interval of measurement uncertainty has limits $y \pm t_{\nu_{\rm eff},0.95} \times \surd{\sum_i w_i}$ with $\nu_{\rm eff}=\left(\sum_i w_i\right)^2/\sum_i \{w_i^2/(n_i-1)\}$.

\section{Conclusion}

This paper has considered the problem of extending the procedure of Type A uncertainty evaluation described in the {\em Guide} \cite{GUM} to accommodate the measurement of vector quantities when there is correlation between the measurements of input quantities. It has presented an appropriate generalization of the Welch-Satterthwaite formula for the calculation of an effective number of degrees of freedom. Results for independently measured vectors \cite{classical} and for scalars measured with dependent errors \cite{generalize} have been combined into a single explicit general procedure that reduces to the procedure of the {\em Guide} in the standard case. The procedure is applicable when a vector measurand of dimension $p$ is a known function of several input quantities of dimension $q\ge p$.

A special case that admits some simplification is where the measurand and the input quantities are complex, as is often the case in radio-frequency and microwave metrology. When used to generate 95\% regions of expanded uncertainty in simulated measurements that involved five complex input quantities, the procedure had a success rate of at least 95\% at each setting of the unknown parameters, as required, while not giving regions as large as would be obtained using the method of the second supplement to the {\em Guide} \cite{supp2}, where that method is applicable.

The direction in which the procedure of the {\em Guide} is extended differs considerably from the direction taken in the first two supplements to the {\em Guide}, where a non-classical approach to statistical inference is taken. It is reassuring to see that the now-familiar methods of the {\em Guide} can be extended to multidimensional problems, and it is important to recognise that the resulting procedures generally have better performance than the methods proposed in the supplements.

\section*{Acknowledgement}

This work was funded by the New Zealand Government as part of a contract for the provision of national measurement standards.

\appendix
\section{Data-processing software}
The methods of evaluating effective degrees of freedom described in this paper are incorporated in a software tool called the {\it GUM Tree Calculator}\/ (GTC), which automates most of the steps required to process data with full uncertainty propagation.\footnote{GTC may be obtained from {\tt http://mst.irl.cri.nz}.} GTC uses a software entity called an {\it uncertain number}\/ to represent a measurement result or, equivalently, a quantity estimate with some associated uncertainty \cite{hall:2006,hall:2012}. Basic mathematical operations are defined that manipulate uncertain numbers and hide the more complicated series of operations required to propagate uncertainty. In this appendix we show how the examples in Sections~\ref{example1sec} and \ref{example2sec} can be handled using GTC.

\subsection{Example 1}
The following GTC code calculates $\widehat{\Gamma}$, obtaining at the same time a covariance matrix and a number of degrees of freedom.
\begin{verbatim}
# The observations
s11 = [
    0.0242-0.0101j,-0.0023+0.2229j,
    0.0599+0.0601j, 0.0433+0.2100j,
    -0.0026+0.0627j
]
g_prime = [
    0.1648-0.0250j, 0.1568-0.0179j,
    0.1598-0.1367j, 0.1198-0.0045j,
    0.3162-0.1310j
]

# The uncertain numbers
S11 = type_a.estimate(
    s11, label='S11'
)
G_prime = type_a.estimate(
    g_prime, label='G_prime'
)

# The result
G = G_prime - S11
G.label = 'G'
\end{verbatim}

The first step defines a pair of `sequences' {\tt s11} and {\tt g\_prime} containing the observations associated with $S_{11}$ and $\Gamma^\prime$.

The next step uses the function {\tt type\_a.estimate} to define the uncertain numbers {\tt S11} and {\tt G\_prime} that are
associated with $\widehat{S}_{11}$ and $\widehat{\Gamma}^\prime$. These software entities encapsulate information about the data, including the estimate, the uncertainties and the degrees of freedom. We can display a summary of this information using the commands

\begin{verbatim}
print summary(S11)
print summary(G_prime)
\end{verbatim}

\noindent
giving
\begin{verbatim}
S11:
    (0.024+0.109j),
    u=[0.012,0.046],
    r=-0.14,
    df=4.0
G_prime:
    (0.183-0.063j),
    u=[0.034,0.029],
    r=-0.65,
    df=4.0
\end{verbatim}
The number in parentheses is the complex estimate, the numbers in square brackets are the standard uncertainties associated with the real and imaginary components of this estimate, {\tt r} is the correlation coefficient between the real and imaginary component estimates and {\tt df} is the number of degrees of freedom.

The third step obtains an estimate of $\Gamma$ using equation~(\ref{hatGamma1}). The uncertain number {\tt G} is associated with $\widehat{\Gamma}$. In GTC, the subtraction of {\tt S11} from {\tt G\_prime} implies not only subtraction of the estimates but also propagation of the associated uncertainty and degrees of freedom information. The results can be displayed using the following commands
\begin{verbatim}
print "Gamma = ", value(G)
print
v_y = variance(G)
print "v_y[1,1] = %G" % v_y.rr
print "v_y[2,1] = %G" % v_y.ir
print "v_y[2,2] = %G" % v_y.ii
print
print "degrees-of-freedom = %G" % dof(G)
\end{verbatim}
\noindent
which produces the following results that can be compared with those given in Section~\ref{example1sec}:
\begin{verbatim}
Gamma = (0.15898-0.17214j)

v_y[1,1] = 0.00131753
v_y[2,1] = -0.000725623
v_y[2,2] = 0.00294249

degrees-of-freedom = 6.85323
\end{verbatim}

\subsection{Example 2}
The following GTC code calculates $\widehat{\Gamma}$, obtaining at the same time a covariance matrix and a number of degrees of freedom.

As in Example~1, the first step is to define sequences containing the observations.
\begin{verbatim}
# The observations
g_prime = [
    -0.220+0.008j, -0.217-0.010j,
    -0.222-0.003j, -0.207-0.018j,
    -0.219-0.009j
]
s11 = [
    -0.072+0.066j, -0.075+0.073j,
    -0.087+0.069j, -0.087+0.092j,
    -0.064+0.071j, -0.079+0.067j,
    -0.069+0.072j
]
s12 = [
    0.112+0.903j,  0.115+0.891j,
    0.094+0.888j,  0.098+0.898j,
    0.092+0.899j,  0.072+0.892j,
    0.102+0.906j
]
s21 = [
    0.106+0.900j, 0.098+0.900j,
    0.085+0.893j, 0.106+0.886j,
    0.114+0.907j, 0.089+0.882j,
    0.108+0.893j
]
s22 = [
    0.044+0.106j,  0.029+0.091j,
    0.019+0.096j,  0.036+0.099j,
    0.026+0.102j,  0.026+0.088j,
    0.022+0.096j
]
\end{verbatim}

The next step is to obtain estimates of the four $S$-parameters and $\Gamma^\prime$ using the function {\tt type\_a.multi\_estimate\_complex}. This function treats the sequences {\tt s11}, {\tt s12}, {\tt s21} and {\tt s22} as simultaneous observations of the components of a multivariate quantity. It returns a sequence of four uncertain numbers {\tt S11}, {\tt S12}, {\tt S21} and {\tt S22} associated with $\widehat{S}_{11}$, $\widehat{S}_{12}$, $\widehat{S}_{21}$ and $\widehat{S}_{22}$ respectively.
\begin{verbatim}
# The uncertain numbers
sp = type_a.multi_estimate_complex(
    (s11,s12,s21,s22),
    labels =('S11','S12','S21','S22')
)
S11, S12, S21, S22 = sp

G_prime = type_a.estimate(
    g_prime,
    label='G_prime'
)
\end{verbatim}

The third step is to estimate $\Gamma$ using equation~(\ref{hatGammaEx2}). Again, propagation of information about the uncertainty and degrees of freedom is implied.
\begin{verbatim}
# The result
num = G_prime - S11
den = S12*S21 + S22*num
G = num/den
G.label = 'G'
\end{verbatim}

\noindent
This gives
\begin{verbatim}
Gamma =  (0.15157+0.13165j)

v_y[1,1] = 3.03498E-05
v_y[2,1] = -2.58341E-05
v_y[2,2] = 4.19894E-05

degrees-of-freedom: 9.0096
\end{verbatim}
These figures can be compared with those given in Section~\ref{example2sec}.

\pagebreak
\addtocounter{table}{-3}

%%% TABLE 2
\begin{table}[h]
\caption{Results for $|\Gamma|=0.2$. Performances of the methods with $n=4$ (top), $n=8$ (middle) and $n=16$ (bottom). $p_i$ is the percentage of successes with method $i$ in $10^4$ trials. $\bar{A}_i/\bar{A}_j$ is the ratio of the mean area of the confidence region with method $i$ to the mean area of the confidence region of method $j$ in $10^4$ trials}
{\small
\begin{center}
\begin{tabular}{rlcccccccccccccccccc}
&&&\multicolumn{2}{c}{${p}_1$}&&\multicolumn{2}{c}{${p}_2$}&&\multicolumn{2}{c}{${p}_3$}
&&\multicolumn{2}{c}{$\bar{A}_1/\bar{A}_2$}&&\multicolumn{2}{c}{$\bar{A}_1/\bar{A}_3$}&&\multicolumn{2}{c}{$\bar{A}_2/\bar{A}_3$}\\
$\rho$&$\kappa$&&min&max&&min&max&&min&max&&min&max&&min&max&&min&max\\
\hline
0 & 0 && 96.3 & 97.1 &&{96.4 }& 97.3 &&{95.1 }& 96.6 && 1.06 & 1.08 && 1.23 & 1.68 && 1.16 & 1.56 \\[2mm]
0.3 & 0 && 96.3 & 97.1 &&{96.4 }& 97.4 &&{\bf 94.6 }& 96.8 && 1.06 & 1.08 && 1.19 & 1.63 && 1.11 & 1.51 \\
0.6 & 0 && 96.2 & 97.2 &&{96.5 }& 97.6 &&{\bf 93.2 }& 97.1 && 1.05 & 1.08 && 1.04 & 1.43 && 0.99 & 1.34 \\
0.9 & 0 && 96.2 & 97.1 &&{97.2 }& 98.0 &&{\bf 90.3 }& 97.3 && 1.01 & 1.05 && 0.63 & 0.90 && 0.61 & 0.88 \\
-0.3 & 0 && 96.3 & 97.1 &&{96.4 }& 97.5 &&{\bf 94.3 }& 96.9 && 1.06 & 1.08 && 1.20 & 1.63 && 1.13 & 1.51 \\
-0.6 & 0 && 96.2 & 97.2 &&{96.6 }& 97.7 &&{\bf 93.1 }& 96.8 && 1.05 & 1.07 && 1.04 & 1.42 && 0.99 & 1.34 \\
-0.9 & 0 && 96.3 & 97.1 &&{97.1 }& 98.0 &&{\bf 90.0 }& 97.3 && 1.01 & 1.05 && 0.63 & 0.90 && 0.60 & 0.87 \\[2mm]
0.3 & 0.15 && 96.3 & 97.2 &&{\bf 96.4 }& 97.4 &&{\bf 94.3 }& 96.6 && 1.06 & 1.09 && 1.20 & 1.64 && 1.12 & 1.52 \\
0.3 & 0.3   && 96.2 & 97.2 &&{\bf 96.4 }& 97.3 &&{\bf 94.3 }& 96.8 && 1.07 & 1.09 && 1.21 & 1.65 && 1.12 & 1.52 \\
0.6 & 0.3   && 96.3 & 97.2 &&{\bf 96.6 }& 97.5 &&{\bf 92.9 }& 96.9 && 1.06 & 1.09 && 1.07 & 1.44 && 0.99 & 1.34 \\
0.6 & 0.6   && 96.2 & 97.2 &&{\bf 96.4 }& 97.3 &&{\bf 92.4 }& 96.6 && 1.07 & 1.12 && 1.09 & 1.48 && 1.00 & 1.35 \\
0.9 & 0.45 && 96.2 & 97.1 &&{\bf 96.8 }& 97.6 &&{\bf 90.3 }& 97.1 && 1.05 & 1.09 && 0.66 & 0.94 && 0.61 & 0.88 \\
0.9 & 0.9   && 96.2 & 97.2 &&{\bf 96.3 }& 97.2 &&{\bf 89.6 }& 96.7 && 1.03 & 1.11 && 0.68 & 0.94 && 0.62 & 0.89 \\
\hline
\end{tabular}

\vskip 3mm

\begin{tabular}{rlcccccccccccccccccc}
&&&\multicolumn{2}{c}{${p}_1$}&&\multicolumn{2}{c}{${p}_2$}&&\multicolumn{2}{c}{${p}_3$}
&&\multicolumn{2}{c}{$\bar{A}_1/\bar{A}_2$}&&\multicolumn{2}{c}{$\bar{A}_1/\bar{A}_3$}&&\multicolumn{2}{c}{$\bar{A}_2/\bar{A}_3$}\\
$\rho$&$\kappa$&&min&max&&min&max&&min&max&&min&max&&min&max&&min&max\\
\hline
0 & 0         && 95.0 & 95.9 &&{95.0 }& 96.1 &&{94.8 }      & 96.1 && 1.01 & 1.02 && 0.86 & 1.09 && 0.84 & 1.07 \\[2mm]
0.3 & 0      && 94.8 & 96.0 &&{95.0 }& 96.1 &&{\bf 93.9 }& 96.3 && 1.01 & 1.02 && 0.82 & 1.05 && 0.81 & 1.03 \\
0.6 & 0      && 94.8 & 96.0 &&{95.1 }& 96.2 &&{\bf 92.3 }& 96.6 && 1.01 & 1.02 && 0.70 & 0.89 && 0.70 & 0.88 \\
0.9 & 0      && 94.6 & 95.9 &&{95.5 }& 96.6 &&{\bf 89.6 }& 97.1 && 0.99 & 1.01 && 0.44 & 0.54 && 0.44 & 0.54 \\
-0.3 & 0     && 94.8 & 95.9 &&{95.0 }& 96.0 &&{\bf 93.9 }& 96.4 && 1.01 & 1.02 && 0.82 & 1.05 && 0.81 & 1.03 \\
-0.6 & 0     && 94.8 & 96.0 &&{95.1 }& 96.2 &&{\bf 92.2 }& 96.6 && 1.01 & 1.02 && 0.70 & 0.89 && 0.70 & 0.88 \\
-0.9 & 0     && 94.8 & 96.0 &&{95.6 }& 96.6 &&{\bf 89.6 }& 96.8 && 0.99 & 1.01 && 0.44 & 0.54 && 0.44 & 0.54 \\[2mm]
0.3 & 0.15 && 95.0 & 96.0 &&{\bf 94.9 }& 96.0 &&{\bf 93.8 }& 96.3 && 1.02 & 1.03 && 0.82 & 1.06 && 0.81 & 1.03 \\
0.3 & 0.3   && 95.0 & 96.0 &&{\bf 94.8 }& 96.0 &&{\bf 93.7 }& 96.2 && 1.02 & 1.03 && 0.83 & 1.06 && 0.81 & 1.03 \\
0.6 & 0.3   && 95.0 & 95.9 &&{\bf 94.9 }& 96.1 &&{\bf 92.2 }& 96.6 && 1.02 & 1.03 && 0.71 & 0.91 && 0.70 & 0.88 \\
0.6 & 0.6   && 95.0 & 96.0 &&{\bf 94.4 }& 95.6 &&{\bf 91.7 }& 96.3 && 1.02 & 1.04 && 0.72 & 0.91 && 0.70 & 0.88 \\
0.9 & 0.45 && 94.8 & 96.0 &&{\bf 94.7 }& 95.9 &&{\bf 89.4 }& 96.7 && 1.03 & 1.04 && 0.46 & 0.56 && 0.44 & 0.54 \\
0.9 & 0.9   && 94.6 & 96.0 &&{\bf 93.3 }& 94.8 &&{\bf 89.0 }& 96.2 && 1.00 & 1.03 && 0.45 & 0.55 && 0.45 & 0.54 \\
\hline
\end{tabular}

\vskip 3mm

\begin{tabular}{rlcccccccccccccccccc}
&&&\multicolumn{2}{c}{${p}_1$}&&\multicolumn{2}{c}{${p}_2$}&&\multicolumn{2}{c}{${p}_3$}
&&\multicolumn{2}{c}{$\bar{A}_1/\bar{A}_2$}&&\multicolumn{2}{c}{$\bar{A}_1/\bar{A}_3$}&&\multicolumn{2}{c}{$\bar{A}_2/\bar{A}_3$}\\
$\rho$&$\kappa$&&min&max&&min&max&&min&max&&min&max&&min&max&&min&max\\
\hline
0 & 0 && 94.6 & 95.6 &&{94.5 }& 95.6 &&{94.5 }& 95.9 && 1.00 & 1.01 && 0.80 & 1.00 && 0.79 & 0.99 \\[2mm]
0.3 & 0 && 94.5 & 95.7 &&{94.4 }& 95.8 &&{\bf 93.8 }& 96.2 && 1.00 & 1.01 && 0.76 & 0.95 && 0.76 & 0.95 \\
0.6 & 0 && 94.6 & 95.6 &&{94.7 }& 95.8 &&{\bf 92.2 }& 96.5 && 1.00 & 1.01 && 0.65 & 0.81 && 0.65 & 0.81 \\
0.9 & 0 && 94.5 & 95.6 &&{94.9 }& 96.0 &&{\bf 89.6 }& 96.8 && 0.99 & 1.00 && 0.41 & 0.49 && 0.41 & 0.49 \\
-0.3 & 0 && 94.5 & 95.6 &&{94.5 }& 95.9 &&{\bf 93.8 }& 96.3 && 1.00 & 1.01 && 0.76 & 0.95 && 0.76 & 0.95 \\
-0.6 & 0 && 94.6 & 95.8 &&{94.8 }& 95.8 &&{\bf 91.9 }& 96.6 && 1.00 & 1.01 && 0.65 & 0.81 && 0.65 & 0.81 \\
-0.9 & 0 && 94.5 & 95.8 &&{94.9 }& 96.1 &&{\bf 89.5 }& 96.9 && 0.99 & 1.00 && 0.41 & 0.49 && 0.41 & 0.49 \\[2mm]
0.3 & 0.15 && 94.7 & 95.7 &&{\bf 94.6 }& 95.6 &&{\bf 93.5 }& 96.1 && 1.01 & 1.01 && 0.77 & 0.96 && 0.76 & 0.95 \\
0.3 & 0.3 && 94.5 & 95.7 &&{\bf 94.3 }& 95.5 &&{\bf 93.5 }& 96.1 && 1.01 & 1.02 && 0.77 & 0.96 && 0.76 & 0.95 \\
0.6 & 0.3 && 94.6 & 95.6 &&{\bf 94.2 }& 95.4 &&{\bf 91.8 }& 96.3 && 1.01 & 1.02 && 0.66 & 0.82 && 0.65 & 0.81 \\
0.6 & 0.6 && 94.6 & 95.6 &&{\bf 93.9 }& 95.1 &&{\bf 91.3 }& 96.2 && 1.01 & 1.03 && 0.66 & 0.82 && 0.65 & 0.81 \\
0.9 & 0.45 && 94.5 & 95.6 &&{\bf 93.7 }& 94.9 &&{\bf 89.3 }& 96.5 && 1.02 & 1.04 && 0.42 & 0.51 && 0.41 & 0.49 \\
0.9 & 0.9 && 94.6 & 95.6 &&{\bf 92.2 }& 94.0 &&{\bf 89.1 }& 96.0 && 1.00 & 1.02 && 0.41 & 0.50 && 0.41 & 0.49 \\
\hline
\end{tabular}
\end{center}
}
\end{table}

\pagebreak
%%%%TABLE 3
\begin{table}[h]
\caption{Results for $|\Gamma|=0.8$. Performances of the methods with $n=4$ (top), $n=8$ (middle) and $n=16$ (bottom). $p_i$ is the percentage of successes with method $i$ in $10^4$ trials. $\bar{A}_i/\bar{A}_j$ is the ratio of the mean area of the confidence region with method $i$ to the mean area of the confidence region of method $j$ in $10^4$ trials}
{\small
\begin{center}
\begin{tabular}{rlcccccccccccccccccc}
&&&\multicolumn{2}{c}{${p}_1$}&&\multicolumn{2}{c}{${p}_2$}&&\multicolumn{2}{c}{${p}_3$}
&&\multicolumn{2}{c}{$\bar{A}_1/\bar{A}_2$}&&\multicolumn{2}{c}{$\bar{A}_1/\bar{A}_3$}&&\multicolumn{2}{c}{$\bar{A}_2/\bar{A}_3$}\\
$\rho$&$\kappa$&&min&max&&min&max&&min&max&&min&max&&min&max&&min&max\\
\hline
0 & 0 && 96.1 & 97.1 &&{96.1 }& 97.0 &&{95.2 }& 96.5 && 1.81 & 1.92 && 1.93 & 2.24 && 1.05 & 1.18 \\[2mm]
0.3 & 0 && 96.0 & 97.0 &&{96.0 }& 97.1 &&{\bf 94.9 }& 96.4 && 1.79 & 1.91 && 1.91 & 2.20 && 1.04 & 1.17 \\
0.6 & 0 && 96.2 & 97.0 &&{96.3 }& 97.2 &&{\bf 94.1 }& 96.3 && 1.75 & 1.89 && 1.82 & 2.08 && 1.00 & 1.13 \\
0.9 & 0 && 95.3 & 96.7 &&{96.1 }& 97.3 &&{\bf 92.9 }& 96.3 && 1.67 & 1.82 && 1.55 & 1.80 && 0.86 & 1.05 \\
-0.3 & 0 && 96.0 & 97.0 &&{96.1 }& 97.0 &&{\bf 95.0 }& 96.3 && 1.80 & 1.90 && 1.92 & 2.20 && 1.04 & 1.17 \\
-0.6 & 0 && 95.9 & 97.0 &&{96.2 }& 97.3 &&{\bf 94.5 }& 96.4 && 1.76 & 1.89 && 1.82 & 2.08 && 0.99 & 1.13 \\
-0.9 & 0 && 95.6 & 96.6 &&{96.0 }& 97.2 &&{\bf 92.8 }& 96.3 && 1.67 & 1.82 && 1.55 & 1.80 && 0.86 & 1.04 \\[2mm]
0.3 & 0.15 && 96.0 & 97.0 &&{\bf 95.6 }& 96.6 &&{\bf 94.1 }& 95.7 && 1.87 & 2.07 && 1.98 & 2.40 && 1.04 & 1.17 \\
0.3 & 0.3 && 96.2 & 97.3 &&{\bf 94.8 }& 96.0 &&{\bf 93.3 }& 95.2 && 1.90 & 2.24 && 2.05 & 2.57 && 1.04 & 1.17 \\
0.6 & 0.3 && 96.2 & 97.1 &&{\bf 95.1 }& 96.1 &&{\bf 92.6 }& 94.8 && 1.88 & 2.16 && 1.95 & 2.41 && 1.00 & 1.13 \\
0.6 & 0.6 && 96.1 & 97.2 &&{\bf 93.2 }& 94.3 &&{\bf 90.4 }& 92.9 && 1.97 & 2.41 && 2.03 & 2.65 && 1.00 & 1.13 \\
0.9 & 0.45 && 95.8 & 96.8 &&{\bf 94.4 }& 95.7 &&{\bf 90.8 }& 94.4 && 1.85 & 2.14 && 1.73 & 2.14 && 0.86 & 1.05 \\
0.9 & 0.9 && 96.5 & 97.6 &&{\bf 90.8 }& 92.6 &&{\bf 87.0 }& 90.9 && 2.00 & 2.38 && 1.85 & 2.39 && 0.85 & 1.04 \\
\hline
\end{tabular}

\vskip 3mm

\begin{tabular}{rlcccccccccccccccccc}
&&&\multicolumn{2}{c}{${p}_1$}&&\multicolumn{2}{c}{${p}_2$}&&\multicolumn{2}{c}{${p}_3$}
&&\multicolumn{2}{c}{$\bar{A}_1/\bar{A}_2$}&&\multicolumn{2}{c}{$\bar{A}_1/\bar{A}_3$}&&\multicolumn{2}{c}{$\bar{A}_2/\bar{A}_3$}\\
$\rho$&$\kappa$&&min&max&&min&max&&min&max&&min&max&&min&max&&min&max\\
\hline
0 & 0 && 94.8 & 95.9 &&{94.9 }& 96.3 &&{94.7 }& 96.0 && 1.17 & 1.20 && 1.08 & 1.20 && 0.91 & 1.00 \\[2mm]
0.3 & 0 && 94.8 & 95.9 &&{94.7 }& 96.1 &&{\bf 94.2 }& 96.0 && 1.17 & 1.20 && 1.05 & 1.17 && 0.89 & 0.98 \\
0.6 & 0 && 94.6 & 95.8 &&{94.9 }& 96.1 &&{\bf 93.7 }& 95.9 && 1.16 & 1.19 && 0.99 & 1.09 && 0.85 & 0.92 \\
0.9 & 0 && 94.4 & 95.5 &&{94.8 }& 96.1 &&{\bf 92.4 }& 95.7 && 1.14 & 1.18 && 0.87 & 0.92 && 0.75 & 0.80 \\
-0.3 & 0 && 94.7 & 95.9 &&{94.8 }& 96.0 &&{\bf 94.4 }& 95.9 && 1.17 & 1.20 && 1.06 & 1.17 && 0.89 & 0.98 \\
-0.6 & 0 && 94.7 & 95.9 &&{94.9 }& 96.1 &&{\bf 93.6 }& 95.9 && 1.16 & 1.19 && 0.99 & 1.09 && 0.85 & 0.92 \\
-0.9 & 0 && 94.3 & 95.5 &&{94.7 }& 96.0 &&{\bf 92.3 }& 95.8 && 1.14 & 1.18 && 0.87 & 0.92 && 0.75 & 0.80 \\[2mm]
0.3 & 0.15 && 94.8 & 95.9 &&{\bf 94.0 }& 95.3 &&{\bf 93.5 }& 95.2 && 1.20 & 1.27 && 1.08 & 1.24 && 0.89 & 0.98 \\
0.3 & 0.3 && 94.9 & 96.0 &&{\bf 93.1 }& 94.5 &&{\bf 92.6 }& 94.6 && 1.20 & 1.33 && 1.09 & 1.30 && 0.89 & 0.98 \\
0.6 & 0.3 && 94.7 & 95.9 &&{\bf 93.4 }& 94.5 &&{\bf 91.9 }& 94.3 && 1.20 & 1.31 && 1.02 & 1.19 && 0.85 & 0.92 \\
0.6 & 0.6 && 95.0 & 96.0 &&{\bf 91.2 }& 92.9 &&{\bf 90.0 }& 92.7 && 1.18 & 1.36 && 1.02 & 1.26 && 0.85 & 0.93 \\
0.9 & 0.45 && 94.7 & 95.8 &&{\bf 92.6 }& 93.8 &&{\bf 90.0 }& 93.6 && 1.17 & 1.30 && 0.89 & 1.01 && 0.75 & 0.80 \\
0.9 & 0.9 && 95.3 & 96.6 &&{\bf 89.4 }& 90.9 &&{\bf 87.2 }& 90.9 && 1.09 & 1.27 && 0.84 & 1.01 && 0.75 & 0.81 \\
\hline
\end{tabular}

\vskip 3mm

\begin{tabular}{rlcccccccccccccccccc}
&&&\multicolumn{2}{c}{${p}_1$}&&\multicolumn{2}{c}{${p}_2$}&&\multicolumn{2}{c}{${p}_3$}
&&\multicolumn{2}{c}{$\bar{A}_1/\bar{A}_2$}&&\multicolumn{2}{c}{$\bar{A}_1/\bar{A}_3$}&&\multicolumn{2}{c}{$\bar{A}_2/\bar{A}_3$}\\
$\rho$&$\kappa$&&min&max&&min&max&&min&max&&min&max&&min&max&&min&max\\
\hline
0 & 0 && 94.5 & 95.7 &&{94.6 }& 95.8 &&{94.6 }& 95.8 && 1.07 & 1.08 && 0.94 & 1.04 && 0.88 & 0.96 \\[2mm]
0.3 & 0 && 94.5 & 95.5 &&{94.7 }& 95.7 &&{\bf 94.2 }& 95.8 && 1.06 & 1.08 && 0.92 & 1.01 && 0.86 & 0.94 \\
0.6 & 0 && 94.4 & 95.5 &&{94.7 }& 95.7 &&{\bf 93.5 }& 95.6 && 1.06 & 1.08 && 0.87 & 0.94 && 0.82 & 0.88 \\
0.9 & 0 && 94.3 & 95.5 &&{94.5 }& 95.7 &&{\bf 92.1 }& 95.9 && 1.05 & 1.07 && 0.76 & 0.81 && 0.72 & 0.77 \\
-0.3 & 0 && 94.5 & 95.8 &&{94.6 }& 95.8 &&{\bf 94.3 }& 95.8 && 1.07 & 1.08 && 0.92 & 1.02 && 0.86 & 0.94 \\
-0.6 & 0 && 94.5 & 95.6 &&{94.6 }& 95.8 &&{\bf 93.6 }& 95.6 && 1.06 & 1.08 && 0.87 & 0.94 && 0.81 & 0.88 \\
-0.9 & 0 && 94.2 & 95.4 &&{94.7 }& 95.6 &&{\bf 92.6 }& 95.5 && 1.05 & 1.07 && 0.76 & 0.81 && 0.72 & 0.76 \\[2mm]
0.3 & 0.15 && 94.6 & 95.6 &&{\bf 93.8 }& 94.9 &&{\bf 93.4 }& 95.1 && 1.08 & 1.14 && 0.94 & 1.07 && 0.86 & 0.94 \\
0.3 & 0.3 && 94.7 & 95.6 &&{\bf 92.8 }& 94.1 &&{\bf 92.1 }& 94.3 && 1.08 & 1.18 && 0.94 & 1.11 && 0.86 & 0.94 \\
0.6 & 0.3 && 94.6 & 95.8 &&{\bf 92.9 }& 94.2 &&{\bf 91.7 }& 94.3 && 1.08 & 1.16 && 0.88 & 1.02 && 0.81 & 0.88 \\
0.6 & 0.6 && 94.7 & 95.9 &&{\bf 90.9 }& 92.4 &&{\bf 89.7 }& 92.7 && 1.04 & 1.19 && 0.85 & 1.04 && 0.81 & 0.88 \\
0.9 & 0.45 && 94.5 & 95.6 &&{\bf 92.0 }& 93.6 &&{\bf 90.4 }& 93.6 && 1.06 & 1.16 && 0.77 & 0.87 && 0.72 & 0.77 \\
0.9 & 0.9 && 94.9 & 95.9 &&{\bf 89.1 }& 91.0 &&{\bf 87.9 }& 91.0 && 0.95 & 1.10 && 0.69 & 0.82 && 0.72 & 0.77 \\
\hline
\end{tabular}
\end{center}
}
\end{table}


\begin{thebibliography}{99}
\bibitem{GUM}1995 {\em Guide to the Expression of Uncertainty in Measurement} (Geneva: International Organization for Standardization)
\bibitem{Welch37} Welch B~L 1937 The significance of the difference between two means when the population variances are unequal.  \newblock {\em Biometrika} {\bf 29} 350--362
\bibitem{Welch47} Welch B~L 1947 The generalization of ``{S}tudent's'' problem when several different population variances are involved.  \newblock {\em Biometrika} {\bf 34} 28--35
\bibitem{Satterthwaite} Satterthwaite F~E 1946 An approximate distribution of estimates of variance components. {\em Biometrics} {\bf 2} 110--114
\bibitem{doe}Willink R 2003 On the interpretation and analysis of a degree-of-equivalence {\em Metrologia} {\bf 40} 9--17
\bibitem{MUAP}Willink R 2013 {\em Measurement Uncertainty and Probability}, Cambridge University Press
\bibitem{pivot}Wang C M, Hannig J and Iyer H K 2012 Pivotal methods in the propagation of distributions {\em Metrologia} {\bf 49} 382--9
\bibitem{ChatfieldCollins}Chatfield C and Collins A~J 1980 {\em Introduction to Multivariate Analysis}, Chapman and Hall
\bibitem{hall04}Hall B D 2004 On the propagation of uncertainty in complex-valued quantities {\it Metrologia}\/ {\bf 41} 173-177
\bibitem{classical}Willink R and Hall B D 2002 A classical method for uncertainty analysis with multidimensional data {\em Metrologia} {\bf 39} 361--9.
\bibitem{GuptaNagar} Gupta A~K and Nagar D~K 2000 {\em Matrix Variate Distributions} (Boca Raton, FL: Chapman \& Hall/CRC Press)
\bibitem{pozar}Pozar D~M 1998 {\em Microwave Engineering} John Wiley and Sons
\bibitem{supp2}Joint Committee for Guides in Metrology 2006 {\em Evaluation of measurement data -- Supplement 2 to the ``Guide to the
expression of uncertainty in measurement'' -- Extension to any number of output quantities}
\bibitem{supp1}Joint Committee for Guides in Metrology 2006 {\em Evaluation of measurement data -- Supplement 1 to the ``Guide to the
expression of uncertainty in measurement'' -- Propagation of distributions using a Monte Carlo method}
\bibitem{MSTpaper}Willink R 2010 Difficulties arising from the representation of the measurand by a probability distribution {\em Measurement Science and Technology} {\bf 21} 015110
\bibitem{Geisser}Geisser S and Cornfield J 1963 Posterior distributions for multivariate normal parameters {\em Journal of the Royal Statistical Society, Series B} {\bf 25} 368--76
\bibitem{hall06}Hall B D 2006 Monte Carlo uncertainty calculations with small-sample estimates of complex quantities {\it Metrologia}\/ {\bf 43} 220-226
\bibitem{generalize}Willink R 2007 A generalization of the Welch-Satterthwaite formula for use with correlated components {\em Metrologia} {\bf 44} 340--9
\bibitem{JohnsonWichern}Johnson~R~A and Wichern~D~W 1998 {\em Applied Multivariate Statistical Analysis} 4th edn.,  Prentice-Hall
\bibitem{hall:2006}Hall B D 2006 Computing uncertainty with {\em uncertain numbers}. {\em Metrologia} {\bf 43} L56 -- L61
\bibitem{hall:2012}Hall B D 2013 Object-oriented software for evaluating measurement uncertainty {\em Meas.\ Sci.\ Tech.} {\bf 24} 055004 (doi:10.1088/0957-0233/24/5/055004)
\end{thebibliography}
\end{document}